\documentclass[article,traditabstract]{aa}

\usepackage{amsmath}
\usepackage{graphicx}
\usepackage{txfonts}
\usepackage[sans]{dsfont}
\usepackage{mathrsfs}
\usepackage{trfsigns}
\usepackage{natbib}
\usepackage[normalem]{ulem}
\usepackage[dvipsnames]{color}
\definecolor{myblue}  {named}{MidnightBlue}
\definecolor{myred}   {named}{RedViolet}
\usepackage[colorlinks=true,
  urlcolor=myred,
  linkcolor=myblue,
  citecolor=myblue,
  bookmarks=true,
  pdfstartview=FitH,
  raiselinks=true,
  pdfborderstyle=U]{hyperref} 
\usepackage{url}
\usepackage{hyperref} 
\usepackage{soul}
\usepackage{footmisc}

\def\thetav{\pmb{\theta}}

\def\yv{\pmb{y}}
\def\muhz{$\mu$Hz}
\newcommand{\argmax}[1]{\underset{#1}{\operatorname{argmax}}}
\newcommand{\argmin}[1]{\underset{#1}{\operatorname{argmin}}}

\def\18sco{18 {Sco}}

\begin{document}

\title{Estimating the p-mode frequencies of the solar twin 18 Sco\thanks{Based on observations collected  at the European Organisation for Astronomical Research in the  Southern Hemisphere, Chile (run ID: 183.D-0729(A))}}
\author{M.~Bazot\inst{1}
\and T.~L.~Campante\inst{1}
\and W.~J.~Chaplin\inst{2}
\and H.~Carfantan\inst{3}
\and T.~R.~Bedding\inst{4}
\and X.~Dumusque\inst{1,5}
\and A.-M.~Broomhall\inst{2}
\and P.~Petit\inst{3}
\and S.~Th\'eado\inst{3}
\and V.~Van Grootel\inst{3}
\and T.~Arentoft\inst{6}
\and M.~Castro\inst{7}
\and J.~Christensen-Dalsgaard\inst{8}
\and Jos\'e-Dias  do Nascimento Jr\inst{7}
\and B.~Dintrans\inst{3}
\and H.~Kjeldsen\inst{8}
\and M.~J.~P.~F.~G.Monteiro\inst{1,9}
\and N.~C.~Santos\inst{1,9}
\and S.~Sousa\inst{1}
\and S.~Vauclair\inst{3}
}

\institute{Centro de Astrof\'{\i}sica da Universidade do Porto, Rua
  das Estrelas, 4150-762, Porto, Portugal; bazot@astro.up.pt
\and
School of Physics and Astronomy, University of Birmingham, Edgbaston,
Birmingham B15 2TT, United-Kingdom
\and 
Institut de Recherche en Astrophysique et Plan\'etologie, Universit\'e de Toulouse UPS--OMP / CNRS, 14, avenue \'Edouard Belin, 31400 Toulouse, France
\and
Sydney Institute for Astronomy, School of Physics, University of Sydney NSW 2006, Australia
\and
Observatoire de Gen\`eve, 51 Chemin des Maillettes, CH-1290, Sauverny, Suisse
\and 
VUC Aarhus, Ingerslevs Boulevard 3, 8100 Aarhus C., Danmark
\and
Depart. de F\'isica Te\'orica e Experimental, Univers.
Federal do Rio Grande do Norte, 59072-970 Natal, R.N., Brazil
\and
Stellar Astrophysics Centre, Department of Physics and Astronomy, Aarhus University, Ny Munkegade 120, DK-8000 Aarhus C, Denmark 
\and
Departamento de F\'{\i}sica e Astronomia, Faculdade de Ciencias, Universidade do Porto, Porto, Portugal
}
\abstract{
Solar twins have been a focus of attention for more than a decade, because their structure is extremely close to that of the Sun. Today, thanks to high-precision spectrometers, it is possible to use asteroseismology to probe their interiors. Our goal is to use time series obtained from the HARPS spectrometer to extract the oscillation frequencies of {\18sco}, the brightest solar twin. We used the tools of spectral analysis to estimate these quantities. We estimate 52 frequencies using an MCMC algorithm. After examination of their probability densities and comparison with results from direct MAP optimization, we obtain a minimal set of 21 reliable modes. The identification of each pulsation mode is straightforwardly accomplished by comparing to the well-established solar pulsation modes. We also derived some basic seismic indicators using these values. These results offer a good basis to start a detailed seismic analysis of {\18sco} using stellar models.
}

\maketitle

\section{Introduction}

In the field of stellar physics, the study of solar twins has recently received growing attention. Since the term was first coined by \citet{CdS81} to designate stars spectroscopically identical to the Sun, they have been the focus of photometric and spectroscopic studies aiming at measuring their fundamental atmospheric parameters \citep{Gustafsson98,Melendez07,Gustafsson08,Melendez10}. On the one hand there has been an on-going race to find the {\textquotedblleft}best{\textquotedblright} solar twin. On the other, samples of such stars have been used to try to answer statistically the question: is the Sun a peculiar star?

Nowadays, thanks to various technical breakthroughs in the field of observational astrophysics, these founding studies can be supplemented by additional measurements. Spectropolarimetric surveys of the solar twins using NARVAL have allowed observers to detect magnetic fields in several objects and to reconstruct their magnetic topology \citep{Petit08}. More recently, \citet[][hereafter Paper~I]{Bazot11} combined interferometric and asteroseismic measurements, from respectively the PAVO beam combiner at the CHARA array and the high-precision spectrograph HARPS at La Silla Observatory, to estimate the linear and acoustic radii of 18~Sco, the brightest solar twin, and to derive its mass. The method used to determine the acoustic radius relied on the use of the autocorrelation of the radial-velocity time series. This paper is the continuation of this study, which now aims at a detailed analysis of the seismic data.

Asteroseismology measures quantities directly sensitive to the stellar interiors. This is an advantage compared to the classical observable quantities, usually obtained from spectroscopy or photometry, which are sensitive to the poorly-modelled external layers of the stars. In contrast, it is possible to extract information from the seismic signal roughly independent of these layers and more robustly described by the existing stellar codes. 

Following the development of high-precision spectrographs (mostly for the search of extrasolar planets) and the space missions CoRoT and Kepler, seismic data have been more and more frequently used to model stars in general, and main sequence sun-like stars in particular \citep[e.g.,][]{Miglio05,Bazot05,Bazot08,Dogan10,Metcalfe10,Brandao11}.
The best-case scenario from the perspective of detailed modelling is to have access to the eigenfrequencies of the stellar pulsation modes and to their characteristic numbers $n,l,m$\footnote{With $n$ the number of nodes of the eigenfunction to the stellar pulsation equations and $l,m$  corresponding to the angular degree and the azimuthal order of the eigenfunction, related to the orthogonal set of spherical harmonics $Y^m_l$, which are solutions to Laplace's equation.}. They are then combined to effectively obtain surface-independent information. In the following, we present the strategy used to determine these frequencies.

In Section~\ref{sect:data} we discuss the data and return to some of the sampling issues evoked in Paper~I. Before proceeding to the frequency analysis we give an overview of the characteristics of the noise affecting the seismic signal in Section~\ref{sect:noise}. We recall the characteristics of the parametric model used to describe the power spectrum in Section~\ref{sect:model}. We define an inverse problem for the estimation of the parameters and cast it into a Bayesian formulation. We then solve it numerically using a Markov Chain Monte Carlo (MCMC) algorithm. Our strategy is to first, test the methodology using simulated time-series and then to apply it to the real data. In Section~\ref{sect:bla}, we discuss several aspects of our results. From a methodological standpoint, we try to assess the robustness of our MCMC strategy by comparing it to another estimation procedure. We also discuss the choice of our parametric model and the priors we used in our Bayesian formulation. From a physical point of view, we measure the impact of our new estimates on the acoustic radius and the stellar mass.

\section{Data}\label{sect:data}

\begin{figure}[t]
\center
\includegraphics[width=\columnwidth]{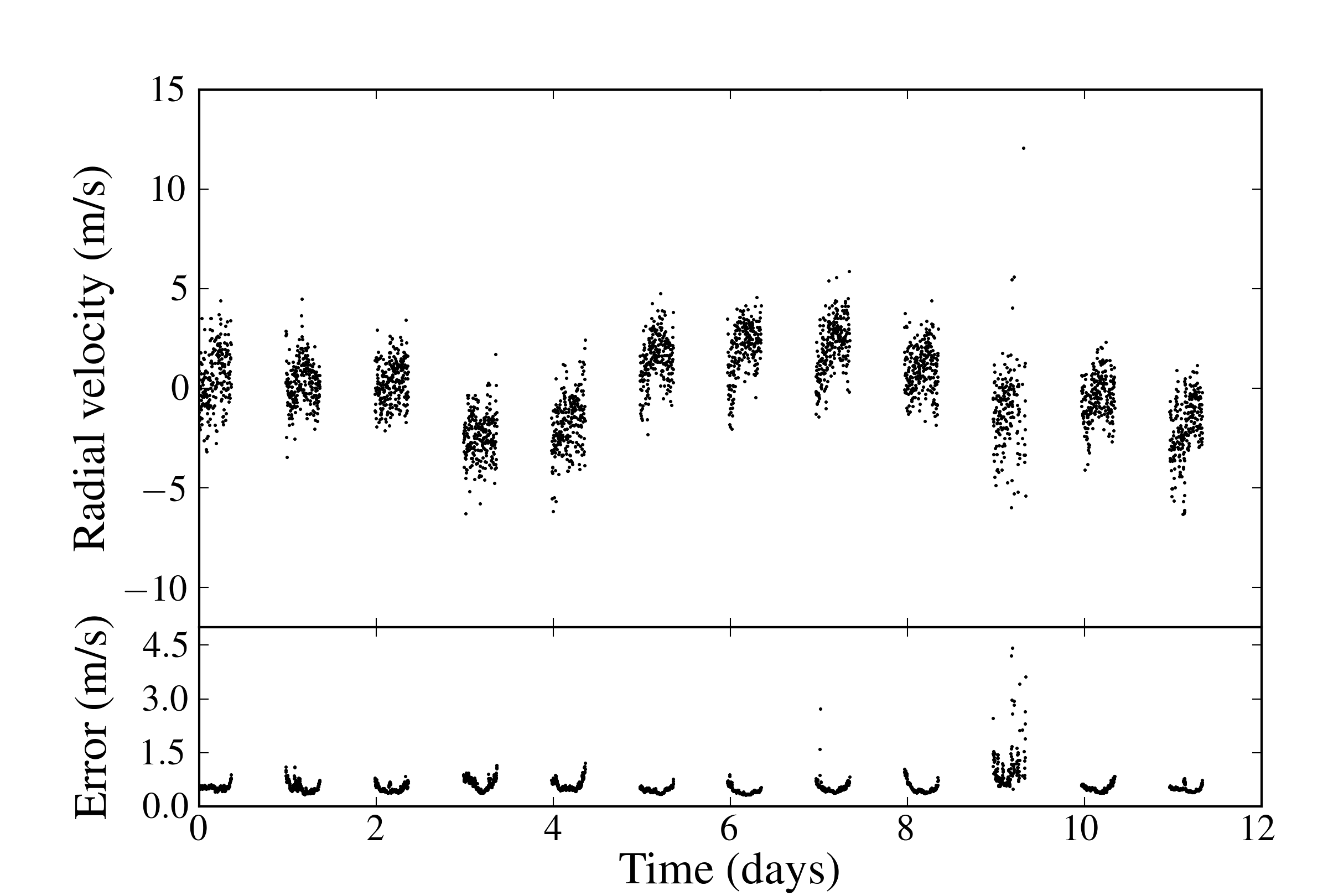}
\caption{Time series of radial velocities (upper panel) and their uncertainties (lower panel) from HARPS observations of 18 Sco. A few points deviating strongly from the bulk of the time series lie outside the plotted range.}
\label{fig:ts}
\end{figure}

The time series was described in Paper~I and is shown in Fig.~\ref{fig:ts} in its unfiltered version. We collected 2833 radial velocity measurements over 12 nights from 10 to 21 May 2009. In this section, we would like to draw the attention on some of the consequences of the window function $w(t) = \sum \delta(t -t_n)$, with $t_n$ the median time of the $n$-th exposure.

The observed signal can be written $y(t) = \tilde{y}(t)w(t)$, with $\tilde{y}(t)$ the continuous time series. When the sampling is uniform, the definition of the Nyquist frequency is $\nu_{\mathrm{N}} = 1/(2\Delta t)$, with $\Delta t$ the sampling time. In the case of unevenly sampled time series, there is, strictly speaking, no Nyquist frequency. However, this does not mean that there is no spectral folding and equivalent frequencies have been suggested \citep[see e.g.][]{Bretthorst00}. 

\begin{figure}[t]
\center
\includegraphics[width=\columnwidth]{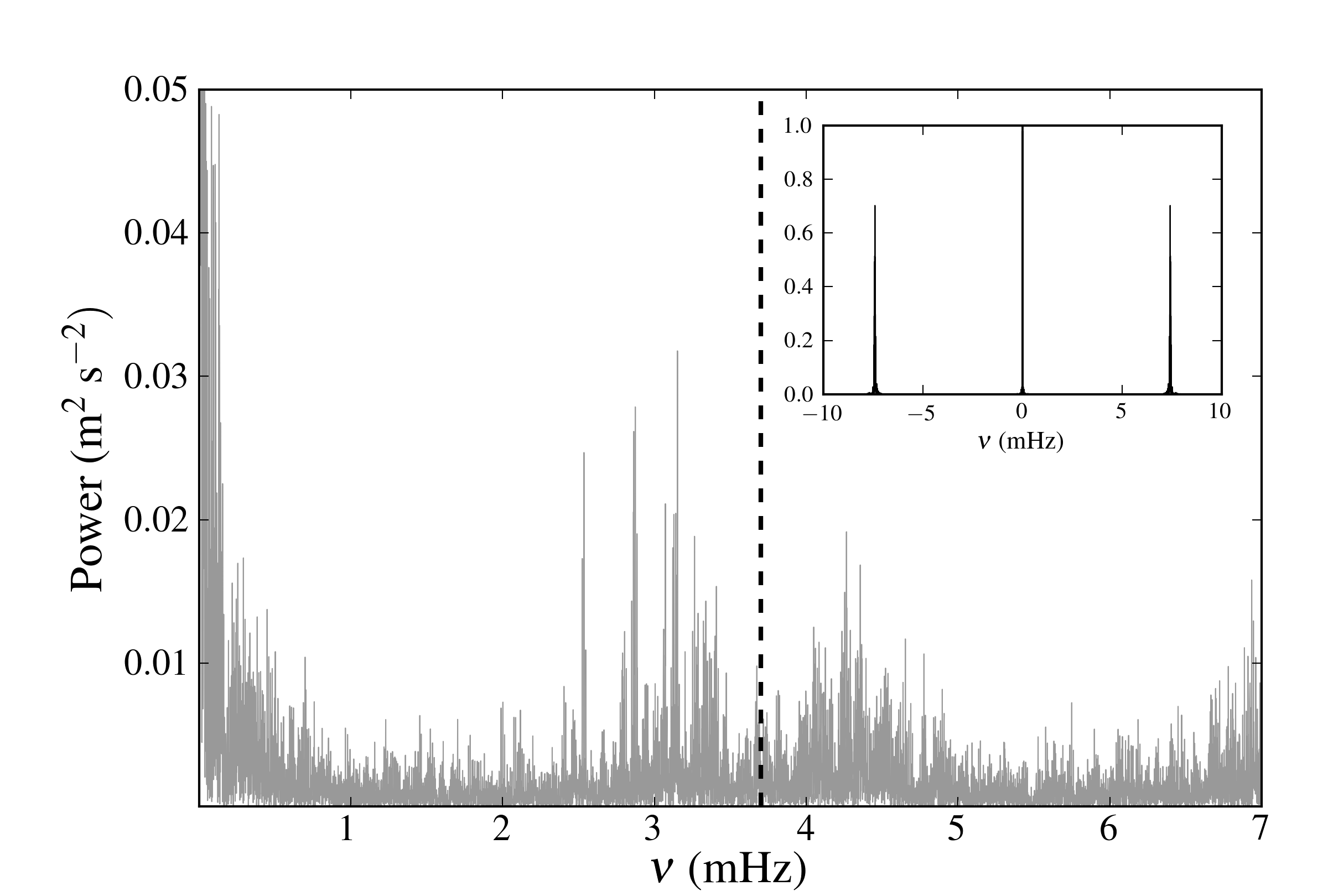}
\caption{Power spectrum of the radial velocity of 18 Sco estimated using a Lomb-Scargle weighted periodogram. The vertical dashed line marks the equivalent Nyquist frequency calculated using the median exposure time. The inset shows the power spectrum of the window function $w(t)$, normalized to its maximum.}
\label{fig:spectrum}
\end{figure}

The power spectrum of {\18sco} is shown in Fig.~\ref{fig:spectrum}. The differences with the figure of Paper~I come from the filtering of the time series\footnote{And from a small numerical bug in the weighting scheme of the LS spectrum in Paper~I. The lower amplitude observed in Fig~\ref{fig:spectrum} should be considered as the reference. This does not affect the results from Paper~I.}. The inset shows the spectral window, that is the squared modulus of the Fourier transform of $w(t)$. Its maximum is at 0 (around which also stand the daily aliases, not visible here, see Paper~I), and two prominent peaks at $\pm7.4$~mHz. They are the aliases caused by the sampling. This means that any signal present at frequencies around $3.7$~mHz is likely to be folded. It should, however, be emphasized that this is not a clear-cut limit, nor an exact estimate; therefore, it is not possible to assess firmly whether some modes at higher frequencies perturb the spectrum below 3.7~Hz or, for that matter, if we are missing some genuine modes above this limit.

It turns out that the median observing time, ${\rm med}(\Delta t)$, offers a reasonably good estimator of the upper limit for folding. This is expected because the sampling is close to uniform. From night to night, the mean values of the observing time lie in the range $132 - 172$~s and their standard deviations within $2 - 232$~s (the upper value resulting mostly from one cloudy night, otherwise, the standard deviations lie in the range $2 - 60$~s, the remaining discrepancy being explained by longer exposures during some of the nights). Furthermore, the median is almost unaffected by the daily gaps. Defining our {\textquotedblleft}equivalent Nyquist frequency{\textquotedblright} as 
\begin{equation}
\nu^{(\mathrm{n.u.})}_\mathrm{N} = \frac{1}{2\mathrm{med}(\Delta t)},
\end{equation}
we obtain an indicative value of 3.7~mHz, in good agreement with the general shape of the spectrum. 

Such a low value of $\nu^{(\mathrm{n.u.})}_\mathrm{N}$ is also a problem when estimating the various noise contributions in the power spectrum. In particular, it makes it difficult to address the question of the photon noise level, which is the main source of noise at high frequencies, i.e. few mHz above the p-mode envelope (see Fig.~\ref{fig:noise}). A common, straightforward approach consists of estimating it by simply calculating an average of the amplitude spectrum well above the p-mode region; this cannot be done here because of the spectral folding.

\section{Noise}\label{sect:noise}

\begin{figure}[t]
\center
\includegraphics[width=\columnwidth]{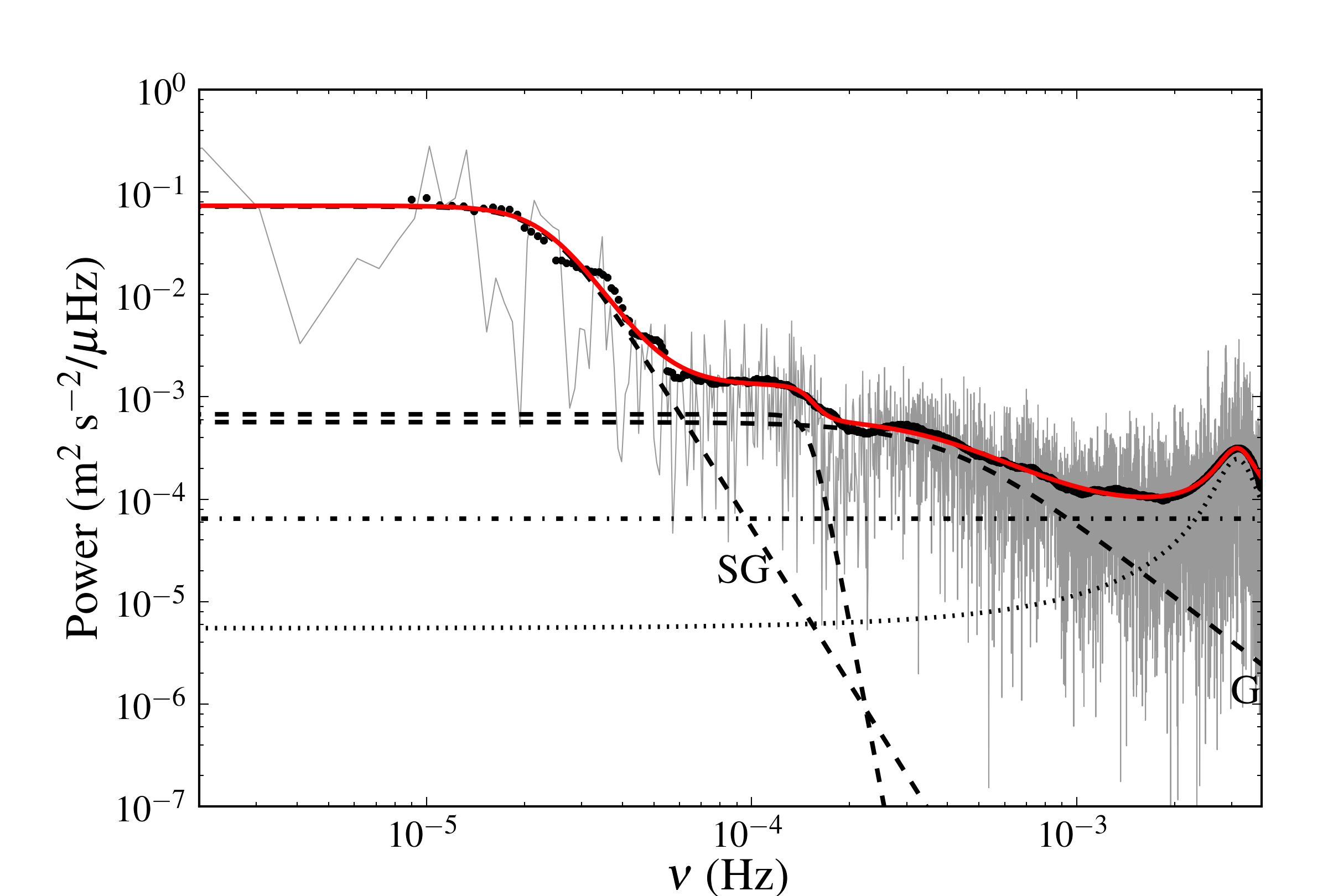}
\caption{Power spectrum of the radial velocity of 18 Sco represented in logarithmic units. The contributions to the noise from granulation (G), supergranulation (SG) and from a third Harvey-like component are represented as dashed lines. The photon noise is a dot-dashed line. The p mode component of the spectrum appears as a dotted line. The full red line shows their sum. The black dots are the points of a binned power spectrum that has been effectively used to perform the fit.}
\label{fig:noise}
\end{figure}

Before considering the extraction of p-mode oscillation frequencies, we discuss briefly the impact of the different noise sources on the relevant signal. It is customary to consider three main phenomena contributing to the overall noise: activity, granulation (at different scales) and instrumental noise \citep[which we assumed to be close to the photon noise, however see][]{Dumusque11}. Characteristic times for activity typically scale as the rotational period. In the case of {\18sco}, a 22.7-d rotation period has been measured from spectropolarimetry \citep{Petit08}. A 12-d time series is certainly not enough to capture this kind of signal. We therefore ignore this contribution.

 \citet{Harvey85} suggested modelling the contributions to the power spectrum coming from granulation as Lorentzian, $4\sigma^2\tau/[1+(2\pi\nu\tau)^2]$, with $\tau$ the characteristic timescale of the process and $\sigma$ the velocity rms. Such functions are representative of non-oscillatory velocity fields whose autocorrelations decay exponentially with time. However, it is difficult to detect this kind of signal precisely enough to assess accurately which of these phenomena contribute effectively to the low-frequency signal. In fact, the exact dependency of these noise components on frequency is unclear and still subject  to discussion \citep[e.g.][]{Guenther08}. Note that it is also possible to use simpler scaling laws of the form $\nu^{-2}$ to model them. In all cases, following \citet{Harvey85}, we should warn that these models are crude.

In seismic studies, it is customary to use parametrized {\textquotedblleft}Harvey functions{\textquotedblright}, $\mathscr{H}(\nu)=\alpha/[1+(2\pi\nu\tau)^{\beta}]$, in order to improve the fit to these low-frequency components. After several trials we found that we needed to use three such functions. The photon noise is assumed to be white. The contribution to the power spectrum of the p modes is modelled as a Lorentzian envelope \citep{Dumusque11}. Figure~\ref{fig:noise} shows the power spectrum corresponding to our time series in logarithmic units. It has been computed using a weighted Lomb-Scargle periodogram \citep{Lomb76,Scargle82,Zechmeister09}. It was evaluated at frequencies separated  by $1/T$, with $T$ the total observing time. To allow comparison with previous studies, it has been normalized according to \citet{Kjeldsen05}. The best fit to the spectrum of this composite model is also shown. For stability, the fit to the background was performed on a heavily smoothed version of the spectrum in order to retain only the slowly-varying features. These values, because of the low {\textquotedblleft}Nyquist{\textquotedblright} frequency, should be considered with care. In particular, it is much easier to perform this fit when the photon-noise constant component can be fixed separately using the signal-free high-frequency regions of the power spectrum\footnote{In fact, the only region of the power spectrum that seems relatively free from stellar granulation noise is the narrow range 1800--2200~$\mu$Hz, where the average noise is 2.6~cm~s$^{-2}$ (against 2.4~cm~s$^{-2}$ for the fitted photon noise).}.

The results of our fit are given in Table~\ref{table:noise}, which displays the parameters of the Harvey functions and the average photon noise. In order to estimate uncertainties on the fitted parameters, we assumed that the binned points in the smoothed power spectrum obey Gaussian statistics, whose variances were evaluated using the rms in each bin. We then used a simple Monte Carlo experiment with 10\,000 simulated spectra to obtain the uncertainties displayed in Table~\ref{table:noise}. 

The Harvey functions are often interpreted as representing different scales of surface convective motions. In the case of granulation, the time scale agrees well with those found in the solar case  by \citet{Harvey85} and \citet{Lefebvre08}. It is possible that the the larger time-scale phenomena correspond to supergranulation, which has been well-studied in the solar case. We however note that the value returned for $\tau$ is lower by one or two orders of magnitude than results found in the by \citet{Harvey85,Title89,DeRosa00} and \citet{Shine00}. It is not possible with such short time series to determine whether it is an intrinsic difference between {\18sco} and the Sun or if this is a methodology-induced effect. The intermediate scale is sometimes interpreted as mesogranulation \citep{Dumusque11}, but its very existence is subject to debate in the solar case \citep[see][and references therein]{Nordlund09}. This debate being outside the scope of this paper, we limit ourselves to mention this intermediate scale without attributing it to surface convection or to an instrumental artifact in the data.

\begin{table}
\caption{Parameters for the best fit to the smoothed spectrum with uncertainties. SG and G stand respectively for the supergranulation and granulation components.}              
\label{table:noise}      
\centering                                      
\begin{tabular}{l c c c}          
\hline\hline
\\[-3.mm]                                   
&$\tau$ (s)&$\alpha$ (m$^2$~s$^{-2}$/$\mu$Hz)&$\beta$\\
\hline
\\[-3.mm]                                   
1 (SG)&$6642 \pm 438$&$(7.2 \pm 1.1)\times10^{-2}$& $5.1\pm 0.3$\\
2     &$1051 \pm 34$ &$(6.8 \pm 0.5)\times10^{-2}$& $16.9\pm 1.3$\\
3 (G) &$389\pm 5$    &$(5.7 \pm 0.1)\times10^{-4}$& $2.47\pm 0.05$\\
\hline
\\[-2.mm]        
&&Power (m$^2$~s$^{-2}$/$\mu$Hz)&\\            
\hline
\\[-3.mm]                                   
Photon noise&&$(6.46 \pm 0.16)\times10^{-5}  $&\\                                                
\hline                                           
\end{tabular}
\end{table}

\section{Modelling of the signal}\label{sect:model}

\subsection{The spectrum model}

In sun-like stars, p modes are excited by turbulent convective motions. Meanwhile, each mode is damped \cite[see e.g][]{Houdek99}. It is assumed that the number of excitation events per damping time is large. The central-limit theorem then ensures that the amplitude distributions of the modes converge towards normal ones \citep{Foglizzo98b}, which translates into a power spectrum that is exponentially distributed. We make the additional assumption that the frequency bins in the Fourier space are independent. In this case, the density probability density of the power spectrum at frequency $\nu_i$ is given by

\begin{equation}
f(p(\nu_i)) = \frac{1}{\mathscr{P}(\nu_i)} \exp\left[{-\frac{p(\nu_i)}{\mathscr{P}(\nu_i)}}\right].
\end{equation}

Ideally, if one obtains many independent measurements of the stellar power spectrum, it will tend, at each $\nu_i$, towards the expectation value $\mathscr{P}(\nu_i)$. The form of this function must be specified from our knowledge of the physical processes governing the mode excitation and the various sources of noise. The expected value of the power spectrum corresponding to one oscillation mode $k$, $\mathscr{P}_k(\nu_i)$, was derived theoretically by \citet{Anderson90}. Considering the equation for a simple damped oscillator with random forcing, they showed that $\mathscr{P}_k(\nu_i)$ has the form of a Lorentzian centred at $\nu_k$, the eigenfrequency of the mode. If we make the additional assumption that the modes are uncorrelated, which is supported by the observations in the solar case \citep{Foglizzo98a}, the expectation value for the multi-modal power spectrum can be written in the form
\begin{equation}\label{eq:sumf}
\displaystyle \mathscr{P}(\nu_i) \simeq \left[\mathscr{P}'(\nu_i)+\mathscr{B}(\nu_i) + \sum_{k=1}^K \frac{V_k^2 H_k}{1+u_{i,k}^2}
  \right]\ast |W(\nu)|^2.
\end{equation}
Here, the $V_k$ represent the visibilities of the modes, $H_k$ their heights, and $u_{i,k} = 2(\nu_i - \nu_k)/\Gamma_k$, with $\Gamma_k$ the linewidth of the Lorentzian. It is related to the mode lifetime through the relation $\Gamma_k = 1/\pi\tau_k$, with $\tau_k$ the lifetime of mode $k$. These parameters are real-valued. $\mathscr{P}'(\nu_i)$ is the power density for all the modes that are not taken into account in the sum \citep[mostly $l=4$ and 5 modes, but also unidentified modes with $l \leq 3$;][]{Fletcher09} but still contribute to the overall power, and $\mathscr{B}(\nu_i)$ the noise contribution, using a $\nu^{-2}$ scaling law to describe the low-frequency component. Note that the convolution by $|W(\nu)|^2$ ($W(\nu)$ being the Fourier transform of the window function) in Eq.~(\ref{eq:sumf}) is only justified because we are considering the expectation value of the power of a stochastic function \citep[][]{Deeming75}, hence assuming our time series are purely stochastic functions. A more rigorous approach can be found in \citet{Stahn08} but was not implemented here.

The model is characterized by a parameter vector $\thetav \in \{\{\pmb{\theta}_k\}_{1\leq k \leq K},\pmb{\theta}_{\mathscr{B}},\pmb{\theta}_{\mathscr{P}'}\}$, which we must to estimate. Here $\pmb{\theta}_k \in \{\nu_k,\Gamma_k,H_k\}$ and has to be estimated for $K$ modes\footnote{This number has to be provided by some ansatz.} (the $V_k$ can be computed straightforwardly provided some assumptions are made on the mode degree). The vectors $\pmb{\theta}_{\mathscr{B}}$ and $\pmb{\theta}_{\mathscr{P}'}$ regroup the parameters describing respectively $\mathscr{B}$ (see Sect.~\ref{sect:noise}) and $\mathscr{P}'$. The inverse problem is now set, as we need to provide estimates of $\pmb{\theta}$ being given $\pmb{y}=(y_1,\dots,y_N)$. This estimation problem, when applied to helio- or asteroseismic data is sometimes referred to as {\textquotedblleft}peak bagging{\textquotedblright}.

A classical approach, first used in helioseismology and then adopted in asteroseismology, involves estimating the parameters of $\mathscr{P}(\nu)$ by minimizing the corresponding negative log-likelihood function \citep{Anderson90}
\begin{equation}\label{likeli}
\displaystyle -\ln(L(\pmb{\theta})) = - \sum_i \ln f(p(\nu_i;\thetav)) = \sum_i \left[\frac{p(\nu_i)}{\mathscr{P}(\nu_i)} + \ln \mathscr{P}(\nu_i)\right],
\end{equation} 
where we used the notation $f(p(\nu_i;\thetav)) = f(p(\nu_i))$ to indicate the dependency on the model parameters\footnote{\label{note:MLE}Using the (frequentist) Maximum Likelihood (ML) method, one defines a Maximum Likelihood Estimator (MLE) as the value of $\thetav$ that minimizes $-L(\theta)$
\begin{equation*}\label{MLE}
\operatorname{MLE}(\pmb{\theta}) = \argmin{\thetav}(-\ln(L(\thetav))).
\end{equation*}}.

\subsection{Bayesian formulation of the problem}\label{sect:bayes}

Studying $L(\thetav)$ can be an challenging task. For likelihoods that are highly non-linear in the parameters, the chance is high that they possess multiple maxima. This makes the identification of high-likelihood regions equivocal and hence complicates the parameter estimation. Furthermore, if the dimension of the space of parameters is large, serious algorithmic problems might occur. In particular, it becomes difficult to even locate the maxima of $L(\thetav)$ because one cannot sample efficiently the space of parameters.

A possible approach to this problem consists of using the Bayes' formula
\begin{equation}\label{eq:bayes}
\pi(\thetav|\yv) \propto \pi(\thetav) L(\thetav),
\end{equation}
where $\pi(\thetav|\yv)$ is the Posterior Probability Density (PPD) of the parameters for fixed data $\yv$ and $\pi(\thetav)$ is the prior density on the parameters. We can therefore explicitly include the knowledge we already have on $\thetav$ in the PPD, so that $\pi(\thetav|\yv)$ becomes the quantity of interest instead of $L(\thetav)$. If the prior is adequately chosen it might significantly restrain the volume to sample in the space of parameters. This potentially offers two advantages: (i) from the statistical point of view, by reducing the number of local maxima, (ii) numerically, some algorithms being more likely to converge when the volume in the parameter space decreases. For general reviews on the Bayesian methodology, we refer to monographs by \citet{Gregory05} or \citet{Robert07}. 

This approach has become very popular in the case of asteroseismology of sun-like stars. The main reason is that realistic priors on the frequency distribution of the pulsation modes can be obtained by using the theoretical asymptotic distribution of the p modes\footnote{\label{fn:asymptotic} The asymptotic relation in the limit of high frequencies (high orders) and low degrees was given to successive orders of approximation by \citet{Vandakurov67} and \citet{Tassoul80} and can be written as $\nu_{n,l} = (n+l/2+\epsilon)\Delta\nu + O(\nu^{-1})$, with $\Delta \nu$ the average large separation, i.e. the inverse of twice the stellar acoustic radius, and $\epsilon$ a phase-related constant.}. 

Bayesian methodology applied to asteroseismology and how it might improve the fitting of the data has been discussed by, e.g., \citet{Brewer07} and \citet{Gaulme09}. In the following, we will mostly focus on the specific setups used in our algorithms, with a particular emphasis on the priors included in our probabilistic models.

\subsection{Markov Chain Monte Carlo estimation}

A widespread approach to Bayesian estimation consists in using Markov Chain Monte Carlo (MCMC) algorithms to approximate the PPD of the parameters. 

The (stochastic) sampling in the parameter space relies on the convergence properties of Markov Chains which, under certain circumstances, generate realizations of a random variable according to a stationary distribution. The underlying idea of MCMC algorithms is thus to produce a Markov Chains whose stationary distribution is the target PPD \citep{Robert99}. MCMC algorithms are a class of numerical methods that aim at approximating PPDs. One of their features is that one can sample a complex distribution using much simpler ones (often called instrumental laws).

The base used here is a Metropolis-Hastings algorithm \citep{Metropolis53,Hastings70}. Because the complexity of the sampling increases with the dimension of the parameter space, an additional scheme was added in order to set the instrumental law used for the sampling. This was done in a burn-in sequence, during which the chain converges towards its stationary distribution\footnote{This burn-in sequence is then discarded when estimating the PPD.}. The algorithm also allows one to run multiple parallel chains in order to use tempering, which is useful to prevent the Markov Chains becoming stuck in local minima. A complete description of the algorithm was given by \citet{Handberg11}.

\subsubsection{Priors and algorithmic setup}\label{sect:MCMCsetup}

The prior density on the frequency is certainly the most characteristic feature of asteroseismology of sun-like star. We first assume that the components $\thetav_{\mathscr{B}}$ and $\thetav_{\mathscr{P}'}$ of $\thetav$ are independent from the other parameters. We fix them before the estimation of the rest of the parameters. This assumption might affect the estimation of the frequency of some low-amplitude mode. However, it improves significantly the convergence of our algorithm. We can reduce the prior probability density to 
\begin{equation}{\label{eq:prior1}}
\pi(\thetav) = \pi(\pmb{\nu},\pmb{\Gamma},\pmb{H}),
\end{equation}
where we used the simplifying notation $\pmb{\nu}=(\nu_1,\dots,\nu_K)$, $\pmb{\Gamma}=(\Gamma_1,\dots,\Gamma_k)$ and $\pmb{H}=(H_1,\dots,H_k)$. 

In practice, we recast the problem of estimating the $K$ linewidths $\Gamma_k$ to the estimation of only two parameters by assuming a linear dependency of the linewidth on the frequency. We considered only the values of the linewidth $\Gamma_{1}$ and  $\Gamma_{2}$ at 2800~$\mu$Hz and 3600~$\mu$Hz respectively. From theoretical modelling and observational results in the Sun, it is clear that this is an oversimplification. However, it is difficult to suggest a proper empirical law and retaining a linear model should at least capture the overall increasing trend of the linewidth with frequency \citep{Houdek99}, although it might not be even be strictly monotonic \citep{Chaplin05}. The relevant parameter space is now $\{\pmb{\nu},\Gamma_1,\Gamma_2,\pmb{H}\}$. Assuming that the $\nu_k$, $\Gamma_1$ and $\Gamma_2$ are all independent parameters, we can write the second term in (\ref{eq:prior1}) as
\begin{equation}{\label{eq:prior2}}
\displaystyle
\begin{split}
\pi(\pmb{\nu},\pmb{\Gamma},\pmb{H})& = \pi(\pmb{H}|\pmb{\nu},\Gamma_1,\Gamma_2)\pi(\pmb{\nu},\Gamma_1,\Gamma_2)\\
& = \pi(\pmb{H}|\pmb{\nu},\Gamma_1,\Gamma_2)\prod_{k=1}^{K}\pi(\nu_k)\prod_{i=1}^{2}\pi(\Gamma_i).
\end{split}
\end{equation}
%where we have used the notation $\pi_X(x)$, X being a random variable (or vector), to differentiate the distributions from the initial prior. 

Mode heights have been fixed as follows. We first smoothed the power spectrum\footnote{This procedure might, of course, depend slightly on the filter used. We neglect this contribution.}, according to \citet{Kjeldsen08}. To each $\nu_k$ we can thus associate a value for the amplitude, $A_k$, which is then converted to height using the relation $H_k = 2A_k^2/(\pi\Gamma_k)$ \citep{Chaplin08}. The first term in the right-hand side of Eq.~(\ref{eq:prior2}) is therefore
\begin{equation}
\displaystyle \pi(\pmb{H}|\pmb{\nu},\Gamma_1,\Gamma_2) = \prod_{k=1}^{K}\delta( H_k - \frac{2A_k^2}{\pi\Gamma_k}).
\end{equation}

For the frequencies and the mode lifetimes we chose uniform distributions\footnote{Such distributions sometimes enter in the quite generic category of uninformative priors. Considering that we severely restrict our individual modes to vary in a small portion of the frequency domain, this seems hardly to apply here\dots} on the regions where the parameters are allowed to vary. The prior density simplifies to
\begin{equation}\label{eq:prior3}
\pi(\thetav) = \prod_{k=1}^{K}\mathds{1}_{a_k}(\nu_k)\delta(H_k - \frac{2A_k^2}{\pi\Gamma_k})\prod_{i=1}^2\mathds{1}_{b_i}(\Gamma_i),
\end{equation}
where
\begin{equation}\label{eq:unit}
\mathds{1}_{a}(\nu) = \begin{cases}
1& \text{if $\nu \in a$},\\
0& \text{if $\nu \notin a$.}
\end{cases} 
\end{equation}
 The intervals $a_k$ were chosen to have widths of 12~$\mu$Hz and be centred on the initial guesses. These were selected by multiplying by 0.991 the solar frequencies from \citet[][see Sect.~\ref{sect:MCMCid} for a discussion]{Broomhall09}. This ensured that the daily aliases at $\pm11.57$~$\mu$Hz of each modes should not perturb the fit. Moreover, provided that the small separations (see Sect.~\ref{sect:separations}) are large enough, this should also limit the possibilities of {\textquotedblleft}mode swapping{\textquotedblright} for close peaks. We consider frequencies only in the range 1965-3700~$\mu$Hz. The total number of frequencies was fixed at $K=52$. This prior effectively filters out the signal that is not included in windows centred on our first guesses for the frequencies. For the linewidths, we set $b_1 = b_2 = [0,5]$~$\mu$Hz. 

This is essentially the same approach as used for the study of Procyon with the same algorithm in \citet{Bedding10}. One has to keep in mind that these are strong assumptions, especially the one on the mode heights. This has to be taken into consideration when comparing the results to other methodologies in Sect.~\ref{sect:bla}.

The simulation made use of 6 parallel chains for tempering. After removal of a $\sim$250\,000 sample burn-in sequence, we obtained an approximate PPD for $\thetav$ based on $\sim$2\,000\,000 samples.

\subsubsection{Statistical analysis}

The MCMC simulation provides us directly with the marginal probability densities of our parameters\footnote{The marginal PPD of a given parameter is simply the PPD integrated over all the other \begin{equation*}\displaystyle \pi(\theta_i|\pmb{y}) = \int \pi(\thetav|\pmb{y}) d\theta_1\dots d\theta_{i-1}d\theta_{i+1}\dots d\theta_N \end{equation*}}. We can thus calculate various statistics that describe these densities. For each of them, we considered the posterior mode and the posterior median. 

The corresponding $100(1-\eta)\%$ credible sets for the modes was defined as the smallest set containing the parameter with probability $1 \geq 1-\eta \geq 0$
\begin{equation}
C^{\mathrm{mode}}_k = \{\nu_k \in a_k | \pi(\nu_k|\yv) \geq q(\eta)\},
\end{equation}
with $q$ the smallest constant such as $P(C_k^{\mathrm{mode}}|\yv) \geq 1 - \eta$. This credible set always include the mode\footnote{The marginal distribution itself was estimated from the MCMC sample using a kernel density estimation algorithm.}.

For the medians, the credible $100\beta\%$ sets were defined as the intervals for which the parameter is higher or lower to these quantities with equal probability, $\beta/2$ with $1 \geq \beta \geq 0$. We computed them using using the cumulative distribution function\footnote{The cdf is obtained directly from the MCMC output} (cdf) 
\begin{equation}
C_k^{\mathrm{median}} = \{\nu_k \in a_k | \frac{1 + \beta}{2} \geq \mathrm{cdf}(\nu_k|\yv) \geq \frac{1 - \beta}{2}\}.
\end{equation}

The corresponding credible intervals are simply the upper and lower values of the corresponding set. We now discuss the relevance of these statistics using our MCMC algorithm on simulated time series.

\subsubsection{Results for simulated time series}\label{sect:mcmc_benchmarking}

Evaluating the performance of estimation methods against artificial data is common practice. Of course, we need simulations realistic enough to ensure the conclusions can be transposed to the real case. 18 Sco offers a good opportunity to carry out such an exercise. Indeed, its similarity to the Sun suggests that we can obtain satisfying mock data. To that effect, we simply need to start from a solar seismic model, which, with respect to asteroseismic standards, is extremely well-known, and scale its parameters accordingly.

In this section, we use only one artificial time series, in particular because of the MCMC algorithm is time-consuming. The artificial time series were constructed using the BiSON solar frequencies. They were multiplied by the ratio of 18 Sco average large separation to the solar one ($\simeq 0.991$). The mode heights and lifetimes were assumed to be solar. These parameters were given as input to the solarFLAG simulator \citep{JR08} to produce evenly sampled time series. These data were subsequently interpolated to the observing times of the 18 Sco data. 

An additional heuristic adjustment has been made in order to produce a realistic data set. We scaled the amplitude of the time series by a constant factor and added to each point the realization of a Gaussian random variable to simulate the photon noise. This was done by setting the noise level in the frequency space and then converting to the time space using the relation $\sigma_{\mathrm{phot}} = N\overline{A(\nu)}/\pi$, with $\sigma_{\mathrm{phot}}$ the scatter due to the photon noise, $N$ the length of the time series, and $\overline{A(\nu)}$ the photon noise level as estimated from the amplitude spectrum. The scaling constant and the variance of the underlying noise distribution were adjusted to reproduce roughly the signal-to-noise ratio observed in the amplitude spectrum of {\18sco}\footnote{Defined for practical purpose as the ratio of the amplitude maximum in the 1500--3700~$\mu$Hz region to the averaged amplitude over the 1800--2200~$\mu$Hz interval.}. We crudely fixed these values to 0.95 and 0.4 cm/s, so that the artificial signal-to-noise ratio is 6.46, compared to the 6.65 value measured for the observations. This gives a maximum amplitude and a noise level in the simulated spectrum within 5\% of the observed ones. 

%\onltab{2}{\input{freq_benchmarking.tex}}
\begin{table*}
\caption{Results from the benchmarking of the estimation algorithms. For the right and left tables, columns 1 and 2 give the orders and degrees of the simulated modes. Column 3 gives the input frequencies to the time series simulator \citep[][only the frequencies we effectively searched for are displayed, more modes with lower amplitude were included in the simulated time series]{JR08}. Column 4 and 5 display the frequencies obtained from our MCMC algorithm using the posterior median (Col.~4) and the mode (Col.~5) estimates alongside with the upper and lower limits of the corresponding credible interval. Column 6 shows the frequencies estimated with the MAP direct-optimization methodology, alongside uncertainties derived by inversing the Hessian matrix.}              % title of Table
\label{table:freq_benchmark}      % is used to refer this table in the text
%\centering                                      % used for centering table
\begin{tabular}{lccccc}          % centered columns (4 columns)
\hline\hline                        % inserts double horizontal lines
$l$ & $n$&  Input & Median& Mode& MAP\\    % table heading
\hline
\\[-3.mm]                                   % inserts single horizontal line
0   &14	   &2074.60	&$2072.33_{-0.33}^{+0.71}$ &$2072.00_{-0.85}^{+1.62}$&$2074.39\pm0.39$\\[0.8mm]
0   &15	   &2208.52	&$2207.75_{-0.13}^{+0.08}$ &$2207.78_{-0.19}^{+0.21}$&$2208.11\pm0.18$\\[0.8mm]
0   &16	   &2341.62	&$2341.60_{-0.51}^{+0.34}$ &$2341.76_{-0.62}^{+0.55}$&$2341.98\pm0.42$\\[0.8mm]
0   &17	   &2473.77	&$2474.77_{-0.62}^{+0.65}$ &$2474.77_{-1.17}^{+1.36}$&$2473.56\pm0.33$\\[0.8mm]
0   &18	   &2606.07	&$2606.27_{-0.47}^{+0.37}$ &$2606.71_{-1.27}^{+0.50}$&$2606.10\pm0.08$\\[0.8mm]
0   &19	   &2739.37	&$2738.42_{-0.33}^{+0.38}$ &$2738.22_{-0.86}^{+1.06}$&$2738.57\pm0.16$\\[0.8mm]
0   &20	   &2872.82	&$2872.56_{-0.57}^{+0.86}$ &$2871.87_{-0.82}^{+2.08}$&$2872.38\pm0.13$\\[0.8mm]
0   &21	   &3006.50	&$3002.07_{-0.57}^{+1.22}$ &$3001.11_{-0.59}^{+1.93}$&$3005.58\pm0.26$\\[0.8mm]
0   &22	   &3140.17	&$3141.63_{-0.33}^{+0.30}$ &$3141.70_{-0.77}^{+0.77}$&$3140.86\pm1.18$\\[0.8mm]
0   &23	   &3273.89	&$3275.11_{-0.36}^{+0.36}$ &$3275.05_{-0.86}^{+0.98}$&$3274.67\pm0.73$\\[0.8mm]
0   &24	   &3408.19	&$3406.99_{-0.41}^{+0.49}$ &$3406.70_{-0.95}^{+1.19}$&$3407.34\pm0.28$\\[0.8mm]
0   &25	   &3542.70	&$3542.89_{-1.05}^{+0.88}$ &$3543.37_{-2.86}^{+1.51}$&$3540.86\pm1.27$\\[0.8mm]
0   &26	   &3677.54	&$3674.49_{-0.72}^{+0.70}$ &$3674.42_{-1.86}^{+1.42}$&$3675.50\pm1.88$\\[0.8mm]
1   &13	   &2002.44	&$2001.50_{-0.04}^{+0.03}$ &$2001.51_{-0.04}^{+0.02}$&$2001.96\pm0.26$\\[0.8mm]
1   &14	   &2137.33	&$2137.00_{-0.46}^{+0.78}$ &$2136.62_{-0.58}^{+1.34}$&$2137.33\pm0.43$\\[0.8mm]
1   &15	   &2271.23	&$2271.32_{-0.47}^{+0.18}$ &$2271.52_{-0.15}^{+0.13}$&$2271.37\pm0.35$\\[0.8mm]
1   &16	   &2403.73	&$2403.16_{-0.21}^{+0.18}$ &$2403.21_{-0.69}^{+0.35}$&$2403.89\pm0.39$\\[0.8mm]
1   &17	   &2536.27	&$2540.48_{-0.37}^{+0.49}$ &$2541.21_{-1.32}^{+0.34}$&$2536.61\pm0.16$\\[0.8mm]
1   &18	   &2669.12	&$2669.39_{-0.09}^{+0.09}$ &$2669.41_{-0.23}^{+0.19}$&$2669.92\pm0.21$\\[0.8mm]
1   &19	   &2802.69	&$2802.13_{-0.11}^{+0.11}$ &$2802.14_{-0.28}^{+0.26}$&$2803.04\pm0.18$\\[0.8mm]
1   &20	   &2936.64	&$2936.45_{-0.13}^{+0.13}$ &$2936.42_{-0.29}^{+0.34}$&$2937.02\pm0.18$\\[0.8mm]
1   &21	   &3070.29	&$3072.04_{-1.64}^{+1.14}$ &$3073.84_{-1.64}^{+1.00}$&$3074.67\pm0.54$\\[0.8mm]
1   &22	   &3204.20	&$3202.73_{-0.28}^{+0.28}$ &$3202.78_{-0.77}^{+0.62}$&$3203.02\pm0.51$\\[0.8mm]
1   &23	   &3338.27	&$3338.60_{-0.21}^{+0.22}$ &$3338.49_{-0.41}^{+0.61}$&$3339.05\pm0.31$\\[0.8mm]
1   &24	   &3472.67	&$3471.80_{-1.12}^{+0.61}$ &$3472.38_{-1.12}^{+0.82}$&$3472.51\pm1.13$\\[0.8mm]
1   &25	   &3607.63	&$3604.87_{-0.91}^{+1.02}$ &$3603.29_{-0.19}^{+0.42}$&$3605.10\pm2.22$\\[0.8mm]
\hline                                             %inserts single line                                                   
\end{tabular}                                                                   \begin{tabular}{lccccc}          % centered columns (4 columns)
\hline\hline                        % inserts double horizontal lines
$l$ & $n$&  Input & Median& Mode& MAP\\    % table heading
\hline
\\[-3.mm]                                   % inserts single horizontal line
2   &13	   &2063.21	&$2061.33_{-0.57}^{+2.20}$ &$2060.73_{-0.20}^{+0.20}$&$2064.18\pm0.43$\\[0.8mm]
2   &14	   &2197.53	&$2196.48_{-0.18}^{+0.09}$ &$2196.54_{-0.19}^{+0.22}$&$2197.01\pm0.14$\\[0.8mm]
2   &15	   &2331.10	&$2332.79_{-1.72}^{+0.55}$ &$2332.96_{-0.25}^{+0.99}$&$2330.55\pm0.34$\\[0.8mm]
2   &16	   &2463.53	&$2463.47_{-0.61}^{+0.52}$ &$2463.57_{-1.39}^{+1.34}$&$2463.76\pm0.53$\\[0.8mm]
2   &17	   &2596.15	&$2595.29_{-0.46}^{+0.30}$ &$2595.64_{-1.18}^{+0.55}$&$2596.00\pm0.08$\\[0.8mm]
2   &18	   &2729.73	&$2726.75_{-0.37}^{+0.29}$ &$2726.94_{-0.89}^{+0.70}$&$2727.57\pm0.17$\\[0.8mm]
2   &19	   &2863.47	&$2861.20_{-0.29}^{+0.33}$ &$2861.00_{-0.65}^{+0.92}$&$2863.65\pm0.15$\\[0.8mm]
2   &20	   &2997.50	&$2996.55_{-0.48}^{+0.40}$ &$2996.97_{-1.39}^{+0.75}$&$2997.59\pm0.24$\\[0.8mm]
2   &21	   &3131.46	&$3131.08_{-0.17}^{+0.16}$ &$3131.11_{-0.42}^{+0.39}$&$3131.55\pm0.39$\\[0.8mm]
2   &22	   &3265.46	&$3263.80_{-0.41}^{+0.35}$ &$3264.08_{-1.11}^{+0.73}$&$3264.55\pm0.52$\\[0.8mm]
2   &23	   &3400.03	&$3398.69_{-0.34}^{+0.31}$ &$3398.88_{-0.89}^{+0.72}$&$3399.34\pm0.46$\\[0.8mm]
2   &24	   &3534.80	&$3530.31_{-0.49}^{+0.65}$ &$3529.53_{-0.58}^{+1.62}$&$3531.10\pm2.02$\\[0.8mm]
2   &25	   &3669.86	&$3667.57_{-1.46}^{+1.58}$ &$3665.18_{-0.83}^{+4.32}$&$3665.32\pm2.00$\\[0.8mm]
3   &12	   &1983.00	&$1984.31_{-1.02}^{+1.11}$ &$1985.79_{-3.45}^{+1.22}$&$1983.74\pm0.48$\\[0.8mm]
3   &13	   &2118.42	&$2115.46_{-0.86}^{+1.19}$ &$2113.83_{-0.88}^{+3.00}$&$2118.57\pm0.38$\\[0.8mm]
3   &14	   &2252.78	&$2251.42_{-1.40}^{+1.66}$ &$2250.00_{-1.63}^{+0.30}$&$2254.28\pm0.68$\\[0.8mm]
3   &15	   &2385.91	&$2385.77_{-0.91}^{+0.86}$ &$2386.00_{-2.27}^{+2.59}$&$2386.54\pm0.82$\\[0.8mm]
3   &16	   &2518.85	&$2519.01_{-0.87}^{+1.00}$ &$2518.67_{-1.32}^{+3.06}$&$2519.77\pm0.18$\\[0.8mm]
3   &17	   &2652.01	&$2649.85_{-0.33}^{+0.34}$ &$2649.73_{-0.91}^{+0.89}$&$2650.13\pm0.22$\\[0.8mm]
3   &18	   &2786.05	&$2789.10_{-0.49}^{+0.53}$ &$2789.09_{-0.92}^{+1.89}$&$2787.33\pm0.23$\\[0.8mm]
3   &19	   &2920.37	&$2923.79_{-0.81}^{+0.60}$ &$2924.46_{-2.03}^{+1.10}$&$2921.05\pm0.47$\\[0.8mm]
3   &20	   &3054.48	&$3054.85_{-2.45}^{+1.28}$ &$3056.51_{-0.95}^{+0.78}$&$3057.32\pm0.62$\\[0.8mm]
3   &21	   &3188.85	&$3187.85_{-1.12}^{+1.21}$ &$3187.58_{-2.87}^{+2.59}$&$3184.63\pm1.26$\\[0.8mm]
3   &22	   &3323.28	&$3324.29_{-1.25}^{+1.31}$ &$3323.05_{-0.98}^{+4.40}$&$3322.36\pm0.46$\\[0.8mm]
3   &23	   &3458.15	&$3457.42_{-1.01}^{+1.92}$ &$3456.20_{-1.30}^{+1.35}$&$3456.01\pm1.20$\\[0.8mm]
3   &24	   &3593.50	&$3595.80_{-1.34}^{+1.09}$ &$3596.28_{-2.22}^{+3.22}$&$3587.87\pm2.49$\\[0.8mm]
\hline                                          
\end{tabular}
\end{table*}

We note that such a photon noise level corresponds to an average amplitude of 2.7~cm/s (or $8.3\times10^{-5}$~m$^2$s$^{-5}$/$\mu$Hz) in the region 1800--2200~$\mu$H. This is similar to the observed value, for which we assumed that only photon noise was present in this interval (Sect.~\ref{sect:noise}). This might be an \emph{a posteriori} warning that this assumption is inadequate. It might also mean that neglecting the term $\mathscr{P}'(\nu)$ in Eq.~(\ref{eq:sumf}) is not valid.
%\begin{flushleft}
\begin{table*}
\caption{Oscillation frequencies detected for 18 Sco along with their radial order $n$ and degree $l$. The quantities given here are the median of the marginal posterior distributions for each eigenfrequency considered in our model. The corresponding credible intervals are given alongside. We use some flags to signal estimates requiring additional caution. (\textasteriskcentered) indicate eigenfrequencies with clear multimodal marginal distributions, (\dag) pulsation modes potentially unresolved, (\textdaggerdbl) pulsation modes that have not passed a hypothesis testing when using the heights estimated with the MAP approach in Sect.~\ref{sect:data_map} and (?) the mode that has not been identified by the MAP approach.}
\label{table:modes_mcmc} 
\centering         
\begin{tabular}{l l@{\hspace{0.3cm}} l@{\hspace{0.3cm}} l@{\hspace{0.3cm}} l} 
\hline\hline                       
\multicolumn{1}{c}{$n$}& \multicolumn{1}{c}{$l=0$}&\multicolumn{1}{c}{$l=1$}&\multicolumn{1}{c}{$l=2$}& \multicolumn{1}{c}{$l=3$}\\    
\hline
$12$    &                                       &                                      &                                       &   $1985.12_{-1.12}^{+0.87}$ (\textdaggerdbl)     \\[.1cm]
$13$    &                                       &   $2001.51_{-0.02}^{+0.02}$ (\dag)      & $ 2062.78_{-0.20}^{+0.03}$ (\dag)       &   $2117.89_{-0.99}^{+0.60}$ (\textasteriskcentered)  \\[.1cm]
$14$    &$2074.41_{-0.40}^{+1.05}$ (\dag)           &   $2140.38_{-0.02}^{+0.03}$ (\dag)      & $ 2198.63_{-0.92}^{+0.88}$ (\textasteriskcentered)    &   $2253.00_{-1.64}^{+1.02}$             \\[.1cm]
$15$    &$2208.79_{-0.77}^{+1.44}$ (\textasteriskcentered,\textdaggerdbl)  &   $2271.66_{-1.18}^{+1.12}$ (\textasteriskcentered)  & $ 2331.20_{-1.02}^{+1.57}$ (\textasteriskcentered)    &   $2383.34_{-0.68}^{+1.35}$ (\textasteriskcentered,\textdaggerdbl)\\[.1cm]
$16$    &$2342.00_{-1.10}^{+0.57}$                 &   $2403.96_{-0.20}^{+0.07}$ (\textasteriskcentered)  & $ 2464.27_{-0.17}^{+2.40}$ (\textasteriskcentered)    &   $2518.43_{-0.42}^{+0.49}$             \\[.1cm]
$17$    &$2473.63_{-1.00}^{+1.79}$ (\textasteriskcentered)      &   $2536.58_{-0.04}^{+0.04}$            & $ 2595.27_{-0.31}^{+0.26}$ (\textasteriskcentered,\textdaggerdbl)&   $2650.84_{-0.36}^{+0.35}$             \\[.1cm]
$18$    &$2607.31_{-0.75}^{+0.26}$ (\textdaggerdbl)          &   $2668.72_{-0.14}^{+0.11}$             & $ 2730.13_{-0.24}^{+0.74}$              &   $2786.75_{-0.08}^{+0.07}$             \\[.1cm]
$19$    &$2740.27_{-1.01}^{+0.31}$ (\textasteriskcentered)       &   $2802.67_{-0.08}^{+0.09}$            & $ 2863.20_{-0.73}^{+0.27}$ (\textasteriskcentered)    &   $2924.12_{-0.48}^{+0.31}$             \\[.1cm]
$20$    &$2873.78_{-0.12}^{+1.61}$ (\textasteriskcentered)      &   $2936.02_{-0.29}^{+0.22}$            & $ 2994.18_{-0.48}^{+2.01}$ (\textasteriskcentered)    &   $3055.39_{-0.17}^{+0.17}$ (\textdaggerdbl)       \\[.1cm]
$21$    &$3005.34_{-0.41}^{+4.41}$ (\textasteriskcentered,\textdaggerdbl) &   $3071.49_{-0.11}^{+0.12}$            & $ 3132.89_{-0.07}^{+0.07}$               &   $3187.21_{-1.24}^{+1.23}$ (\textdaggerdbl)      \\[.1cm]
$22$    &$3140.10_{-0.06}^{+0.06}$                &   $3202.06_{-0.19}^{+0.53}$ (\textasteriskcentered)  & $ 3263.98_{-0.17}^{+0.16}$              &   $3322.77_{-0.22}^{+0.14}$              \\[.1cm]
$23$    &$3275.80_{-0.17}^{+0.14}$ (\textasteriskcentered)      &   $3338.91_{-0.14}^{+0.13}$            & $ 3397.52_{-0.17}^{+0.29}$              &   $3458.25_{-1.75}^{+1.68}$ (\textdaggerdbl)       \\[.1cm]
$24$    &$3408.53_{-0.26}^{+0.36}$ (\textasteriskcentered)      &   $3475.34_{-0.53}^{+0.53}$ (\textasteriskcentered)  & $ 3531.08_{-0.96}^{+1.69}$ (?)       &   $3595.93_{-1.01}^{+1.00}$ (\textdaggerdbl)       \\[.1cm]
$25$    &$3545.96_{-1.01}^{+0.44}$ (\textasteriskcentered,\textdaggerdbl) &   $3605.76_{-0.27}^{+0.40}$            & $ 3667.94_{-0.33}^{+0.26}$              &                                       \\[.1cm]     
$26$    &$3674.80_{-0.45}^{+0.53}$                &                                       &                                       &                                      \\[.1cm]     
\hline                             
\end{tabular}
\end{table*}
%\end{flushleft}

The time series was analyzed using the MCMC algorithm set as described above.

 Table~\ref{table:freq_benchmark} gives the results of our numerical experiment. It displays the mode and median estimates of the frequencies and their corresponding credible intervals. We also listed the input frequencies to the time series simulator. These results might guide us as to which summary statistic to use. The median capturing the central tendency of the distribution, it might often lead to a better fit to the data than the mode of the marginal posterior. However, we notice that 35\% of the input frequencies are contained in the credible intervals for the modes, whereas only 30\% of them are in the median credible intervals. The mode credible intervals are in general larger than those computed for the median, and quite often the former encompass the latter. These low values might indicate that we underestimate the uncertainties in both cases. On the other hand, if we define an average absolute error as $\sum_k|\widehat{\nu}_k^{}-\nu_k^{\mathrm{input}}|/K$ with $\widehat{\nu}_k^{}$, $k=1,\dots,K$, the estimated frequencies, we find a larger value for the modes (1.76~$\mu$Hz) than for the medians (1.31~$\mu$Hz). This indicates that the median is indeed, on average, more accurate than the mode.

A closer look at the marginal densities could give some explanation to this. We often observe very skewed or multi-modal marginal posterior densities for the eigenfrequencies of our model. It is especially true at low and high frequency, i.e. for low and high radial orders. The median is a better estimator in $\sim$66\% of the case, almost all corresponding to this kind of difficult situations. Therefore, and considering that the credible intervals are conservative enough for both statistics, we opted for the estimator leading to the most accurate and precise values, i.e. the median. 

\begin{figure}[t]
\center
\includegraphics[width=\columnwidth]{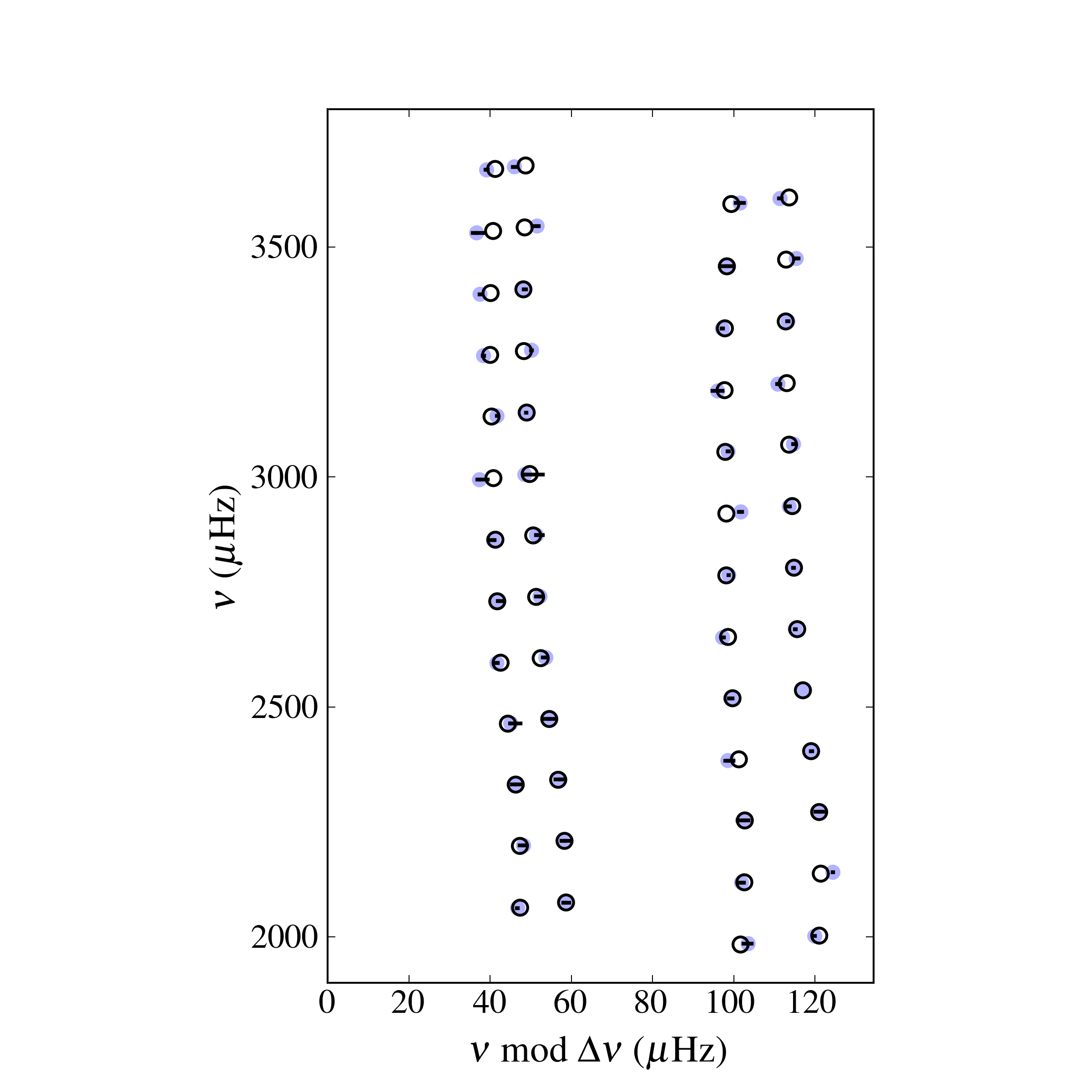}
\caption{Echelle diagram for {\18sco}. The blue dots represent the frequencies given in Table~\ref{table:modes_mcmc}. The $68.3\%$ credible interval are also represented. The black circles show the scaled BiSON frequencies for the Sun \citep[their best-fitting model, Table~1 in][]{Broomhall09}. To plot this \'echelle diagram, we used the value of the large separation given in Paper~I.} 
\label{fig:echelle}
\end{figure}

It should also be noted that, at low frequencies, some densities are very strongly peaked around a central eigenfrequency. This translates into very narrow credible intervals that seem to be unrealistic. This can be explained by the fact that these pulsation modes have long lifetimes and are thus unresolved by our time series. Thus, they appear in the spectrum as Dirac functions and are not properly modelled by a Lorentzian. They should be considered carefully. At the very least the credible intervals associated with these particular frequencies are meaningless.

\subsubsection{Mode identification}\label{sect:MCMCid}

The problem of identification, i.e. associating a couple $(n,l)$ to the frequencies\footnote{Rotation is neglected, therefore $m=0$.} detected in the time series, can be extremely difficult to solve. The standard way to proceed consists in comparing the observations to the theoretical asymptotic formula (see note~\ref{fn:asymptotic}). However, there are many instances of seismic studies, both from space and ground-based, whose outcome led to uncertainty on the identification \citep{Martic04,Carrier03,Kjeldsen05,Appourchaux08,Benomar09,Bedding10}. This of course may be pathological when confronted with single-site ground-based observations. This is problematic from the modelling standpoint, because if one wants to compare observed and theoretical frequencies (or, see Sect.\ref{sect:separations}, rational functions of the $\nu_{n,l}$), one needs to be able to identify precisely the observed modes. 

We are thus in a very particular position when studying 18~Sco, in which we can confirm \emph{a posteriori} the orders and degrees we assigned for each mode. If we make the assumption that in a close vicinity of the Sun the surface related terms in the asymptotic relation do not vary significantly, we can consider that the individual frequencies scale roughly as the average large separation, i.e. as the mean density of the star.  

Such a situation is expected if the stars are close to homologous (which in fact is never really the case for main-sequence stars). This is somewhat justified by the findings of Paper~I, giving a ratio for the large separations
$\Delta \nu_{\odot}/\Delta \nu_{\mathrm{18 Sco}} \sim$ 1.0007. Therefore, we simply assign the $(n,l)$ by comparing directly the frequencies of 18 Sco to the scaled solar ones as suggested by \citet{Bedding10b}. A graphical check is sufficient to do so, as seen in Fig.~\ref{fig:echelle}, in which are plotted the solar modes obtained from BiSON \citep{Broomhall09}.

\subsubsection{Observed frequencies}

We analyzed the observed time series using the method described in Sect.~\ref{sect:MCMCsetup}. For the 52 fitted modes, we found credible intervals consistent with those estimated from the simulated time series. Given the comments made above, we are confident in their robustness. In some cases we obtained very narrow credible sets, whose upper and lower limits, relative to the median value, are $<0.1$~$\mu$Hz in absolute value. These are always high amplitude modes. In Table~\ref{table:modes_mcmc}, we give the medians of the marginal densities alongside the corresponding 68.3\% credible regions. They are shown in Fig.~\ref{fig:echelle}.

The marginal probability densities are qualitatively in good agreement with the results from the simulations. However more of them exhibit multiple maxima, perhaps because of the lower photon noise level used in the simulations. This is always problematic when using the median, since it might bias the estimation towards maxima not corresponding to the real frequency. We flagged these modes in Table~\ref{table:modes_mcmc}. Note, however, that the process remains subjective, and we considered only the densities for which the secondary maxima were obvious. These modes can be used in subsequent studies, but one should keep in mind that the summary statistics we used do not capture all the features of the distribution. The average width of the 68.3\% credible regions on the median is slightly smaller than the one observed for the simulated time series, but this should not prove significant. Compared to the simulations, we also found more skewed distributions. This reinforces our choice to use the median for the statistical summary.

We again observe some very sharply peaked marginal densities. Such shapes lead to very narrow and unrealistic credible regions, for instance much narrower than has been found from space missions with much longer observing baselines \citep[see e.g.][]{Gaulme09, Mathur10, Campante11}. They appear at low frequencies where the mode lifetimes are known to be longer, and we can thus express doubts about the resolution of modes (0,14), (1,13), (1,14), and (2,13), which are flagged accordingly in Table~\ref{table:modes_mcmc}. They are all located at low frequencies, $\nu \lesssim 2150$~{\muhz}. This give an approximate limit above which the modes start to be resolved. In the case of modes (1,13) and (1,14), the kernel estimations of the densities return a numerical error. This might indeed confirm that they cannot be approximated by a continuous density, which would be characteristic of an unresolved pulsation mode. These modes might very well be real. However, because the Lorentzian model becomes incorrect for unresolved modes, the associated uncertainties are likely to be widely underestimated.

Finally, and anticipating the discussion below, some modes, when estimated with an alternative strategy which does not fix the heights (see Sect.~\ref{sect:bla}) did not pass an hypothesis testing. We flagged them and do not recommend their use in subsequent studies. 

The results of our MCMC simulation are available at \href{www.astro.up.pt/~bazot}{\url{www.astro.up.pt/~bazot/data/18ScoII/}}.

\section{Discussion}\label{sect:bla}

\subsection{Comparison to a MAP approach}\label{sect:MAP}

Our goal in this section is to compare the performance of our MCMC approach with another Bayesian strategy based on the direct optimization of the PPD. Note that this is a very general comparison, since not only do we change the \emph{a posteriori} estimators (median and maximum of the PPD), but we also modify our probabilistic model, i.e. the priors. We chiefly want to get an idea of how consistent they might be. This is good procedure to cross-check results obtained using different methodologies. In the case of 18~Sco, given the relatively difficult nature of the data, we see this as necessary.

In the Maximum A Posteriori (MAP) approach, the likelihood is replaced by the PPD. The estimator for the model parameters becomes

\begin{equation}\label{eq:reg_likeli}
\operatorname{MAP}(\thetav) = \argmax{\thetav}(\pi(\thetav|\yv)).
\end{equation}
This estimator is sometimes called the regularized likelihood. This is a common strategy and has been used in the case of the Sun \citep{Chaplin02,Broomhall09} and stars observed from satellites \citep{Appourchaux08,Deheuvels10}. Here, the regularized likelihood is maximized using a Powell algorithm. 

\begin{table*}
\caption{Estimated frequencies of {\18sco} from the MAP method using direct optimization. The associated uncertainties were derived from the inversion of the Hessian matrix.}              % title of Table
\label{table:freq_map}      % is used to refer this table in the text           
\centering         
\begin{tabular}{l l@{\hspace{0.3cm}} l@{\hspace{0.3cm}} l@{\hspace{0.3cm}} l} 
\hline\hline                       
\multicolumn{1}{c}{$n$}& \multicolumn{1}{c}{$l=0$}&\multicolumn{1}{c}{$l=1$}&\multicolumn{1}{c}{$l=2$}& \multicolumn{1}{c}{$l=3$}\\    
\hline
$12$&                &                &                &$1984.32\pm0.68$ \\
$13$&                &$2002.00\pm0.26$&$2063.13\pm0.16$&$2118.01\pm0.10$ \\
$14$&$2074.88\pm0.17$&$2140.60\pm0.12$&$2199.47\pm0.39$&$2254.95\pm0.24$ \\
$15$&$2209.49\pm0.19$&$2270.54\pm0.17$&$2330.74\pm0.20$&$2385.35\pm2.17$ \\
$16$&$2339.14\pm0.18$&$2404.25\pm0.49$&$2467.29\pm0.78$&$2519.03\pm0.44$ \\
$17$&$2475.62\pm0.67$&$2536.84\pm0.11$&$2595.50\pm0.15$&$2651.51\pm1.24$ \\
$18$&$2605.46\pm0.15$&$2669.69\pm0.97$&$2730.41\pm1.06$&$2786.96\pm0.20$ \\
$19$&$2741.34\pm1.54$&$2803.55\pm0.11$&$2863.74\pm0.17$&$2918.07\pm1.40$ \\
$20$&$2874.13\pm0.20$&$2936.92\pm0.69$&$2994.29\pm1.18$&$3055.77\pm0.61$ \\
$21$&$3006.00\pm1.35$&$3071.97\pm0.48$&$3133.10\pm0.43$&$3186.07\pm1.53$ \\
$22$&$3140.58\pm0.40$&$3203.11\pm0.95$&$3264.55\pm0.57$&$3322.89\pm0.40$ \\
$23$&$3276.19\pm0.68$&$3339.44\pm0.37$&$3398.24\pm0.52$&$3459.01\pm2.76$ \\
$24$&$3408.74\pm0.60$&$3476.24\pm1.81$&$3545.86\pm3.58$&$3593.56\pm2.68$ \\
$25$&$3556.15\pm3.92$&$3610.70\pm2.32$&$3669.81\pm2.20$&      \\
$26$&$3679.63\pm2.51$&                &                &      \\
\hline                             
\end{tabular}
\end{table*}

\subsubsection{Setup and tests}\label{sect:MAPsb}
%\onltab{4}{\input{freq_map.tex}}
An interesting feature of this direct optimization approach is that it  converges somewhat more easily than our MCMC algorithm. Therefore, we were able to use less constraining priors. The mode lifetimes and heights were both left free to vary. We applied on the frequencies priors close to those described by Eq.~(\ref{eq:prior3}). The main difference is that, instead of considering individual frequencies, we considered pairs of frequencies of similar parities and differing by one radial order. The $a_k$ were set to $\pm22$~$\mu$Hz above the $l=0$ (or 1) and below the $l=2$ (or 3) modes \citep{Fletcher09}.

It should be noted that the uncertainties in the MAP framework are estimated by inverting the Hessian matrix of the parameters. This is well-justified if the errors, and their second derivatives with respect to the parameters, are small and if these derivatives are not much correlated with the errors. Moreover, the posed fitting problem must also be well-constrained, otherwise the formal uncertainties will also be a poor representation of the true uncertainties. In our case, given the relative complexity of the probability distributions we consider (as can be seen from the MCMC results), these assumptions are not likely to hold. We also note that, from one mode to the other, the Hessian-derived and MCMC uncertainties might differ significantly from the MCMC estimates.

 Another issue with the MAP estimation approach is the potential instability of the minimization algorithm with respect to the initial guesses. To test this, we varied randomly the first-guess frequencies -- using a top-hat distribution of width $\pm 3\,\rm \mu Hz$ -- and the final frequencies were median estimates over 1000 such fits. We found that the scatter in the estimate over these 1000 fits is similar to or larger than the Hessian-derived uncertainties. It was the possibility of deriving more realistic uncertainties that ultimately led us to choose the MCMC estimates as a reference.

We tested our MAP algorithm using the artificial data described in Sect.~\ref{sect:mcmc_benchmarking}. The results are given in Table~\ref{table:freq_benchmark}, alongside those from the MCMC approach. The estimated frequencies are extremely close for the two methods. Nevertheless, the MAP algorithm performs better in terms of accuracy, with the average absolute error being lower than from the MCMC estimates values. 

This opens the door to questions with respect to the proper use of priors in Bayesian analysis. Indeed, if this relative lack of accuracy in the MCMC approach is caused by a bias introduced by the stringent prior constraints imposed, it means that those included in our MAP setup are better. However, this is a difficult problem, numerically, to assume such a great variation of the mode lifetimes and heights in the MCMC approach. This shows how delicate it is to choose between different Bayesian methodologies. Note however that for sufficiently long and precise measurements the two approaches should converge.

\subsubsection{Estimates from the observations}\label{sect:data_map}

Applying the MAP algorithm to the real data, we found results, given in Table~\ref{table:freq_map} for reference, similar to those of the MCMC approach for the frequency estimates. The uncertainties are consistent with our tests using simulated time series. There is only one inconsistency in the identification between the two approaches. The $(l=2,n=24)$ mode was identified as $(l=0,n=25)$ in the MCMC framework. It is likely that the $(l=0,n=25)$ frequency returned from optimization is in fact an alias of the real mode. However, these are very low-amplitude peaks, rejected in the hypothesis testing (see below).

It should also be noted that the greater {\textquotedblleft}flexibility{\textquotedblright} we have in terms of convergence has allowed us to estimate the mode heights and the lifetimes. This allowed us to perform \emph{a posteriori} some hypothesis testing for each mode \citep{Appourchaux09}. It is somewhat more satisfying to carry out such tests on genuinely estimated heights rather than on fixed heights derived from a filtered spectrum, such as the ones used for our MCMC simulations. Therefore, in Table~\ref{table:modes_mcmc}, we flag the values with positive hypothesis testing results. This way, one can chose or not to include them when using the list of frequencies. Note that only 40 frequencies were incompatible with the $H_0$ (null) hypothesis, which tests here the hypothesis that the peak is due to the noise in the data.

\subsection{Comparison with time-domain modelling}

Another approach commonly used in asteroseismology involves representing the signal in the time domain

\begin{equation}\label{eq:sumt}
\displaystyle \tilde{y}(t_n) = \sum_{k=1}^M c_k\sin(2\pi \nu_k t_n) +
d_k\cos(2\pi \nu_k t_n)+\epsilon_n,
\end{equation}
where $(c_k,d_k)$ and $\nu_k$ are the amplitudes and frequencies of
the pulsation modes and $\epsilon_n$, the noise.

Comparing with model (\ref{eq:sumf}), some shortcomings of model (\ref{eq:sumt}) are clear. In particular, it does not take into account the fact that the modes have finite lifetimes. This may lead to an over-fitting of the signal in the vicinity of some modes, i.e. several sine functions being required to reproduce what is actually the dual of a Lorentzian. However, this effect clearly depends on the ratio of the characteristic damping time to the length of the time series. If it is large, then the chances are high that the mode will be unresolved, the power excess largely being confined to one frequency bin (see Sect.~\ref{sect:bla}). In this case, model (\ref{eq:sumt}) will be accurate enough. The unresolved-mode assumption has often been made in the case of ground-based seismic observations \citep{Kjeldsen95,Bouchy01,Bouchy05,Bazot07,Kjeldsen05,Bedding10}. Because of its simplicity, interesting methodologies can be applied to the problem of estimating its parameters. We consider two here. The first one is the well-known CLEAN algorithm \citep{Gray73,Roberts87}, used for iterative deconvolution, and the second one is the {\tt SparSpec} algorithm, a penalization approach to minimization in the context of spectral analysis \citep{Bourgui07}. We used both methods to make sure that the results discussed below are not due to algorithmic artifact.

 The objective of this section is to compare our results with simulations in order to understand the impact of our choice for a physical model (time-domain representation against frequency domain representation) for the power spectrum. We also try to understand, at least qualitatively if this is more important than our choice for the priors on the parameters included in our probabilistic description. We explain how, in the case of {\18sco}, a frequency-domain representation is an improvement for gapped and irregularly sampled time series, for which models such as Eq.~(\ref{eq:sumt}) have previously been used.

\subsubsection{Performances of the methods}

We used a sample of 100 artificial time series constructed as described in Sect.~\ref{sect:mcmc_benchmarking}. They only differ by the realizations of the low-frequency and white noises. We applied the MAP, CLEAN and {\tt SparSpec} algorithms to each element of this sample. The MAP setup is similar to the one described in Sect.~\ref{sect:MAPsb}. In the case of CLEAN, we limited our search to the 1500--3700~$\mu$Hz region. We set a threshold for the detection at three times the noise level in the 1800--2200~$\mu$Hz interval, that is 7.2 cm.s$^{-1}$. The relevant parameter for {\tt SparSpec} is the penalization factor \citep{Bourgui07}, which we empirically set to 0.34, so that the results are close to those obtained for CLEAN. 

Note that we did not include the MCMC algorithm in this comparison. This should not be a problem since we have seen that the results from the direct MAP optimization and the MCMC agree well. Our main goal is to evaluate the efficiency of our algorithm, i.e. how the estimated frequencies reproduce the input frequencies to the time series simulator. We are not concerned with the uncertainties on the parameters here, which were the main reason to retain the MCMC estimates as our reference. Therefore, since we have seen in Sect.~\ref{sect:MAP} that the MAP and MCMC approach lead to close enough estimates, one can extrapolate the following discussion to the MCMC case.

\begin{figure}[t]
\center
\includegraphics[width=\columnwidth]{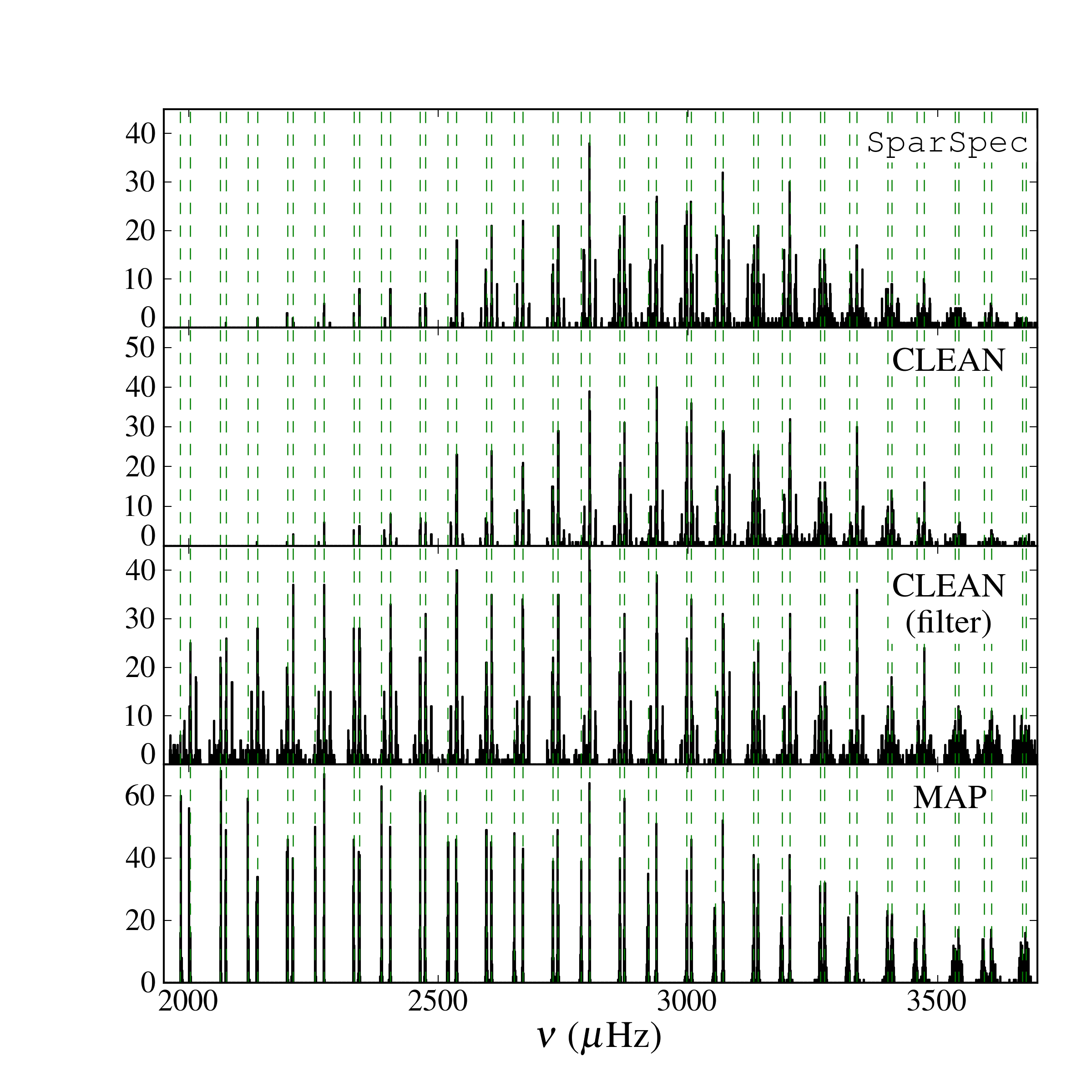}
\caption{Histograms of the output frequencies from the MAP, CLEAN (with and without prior filtering) and {\tt SparSpec} algorithms for 100 artificial time series. The vertical green lines mark the input frequencies to the time-series simulator.}
\label{fig:hist_all}
\end{figure}

In the case of CLEAN and {\tt SparSpec}, it is not possible to go through the normal adjustment of the outputs of the algorithms for the 100 realizations of the time series\footnote{These adjustments consist in removing manually the peaks that are obviously due to noise or aliases not properly removed by the algorithm. In a sense, this is very much similar to applying some {\it prior} knowledge one would have on the frequencies, but after the estimation process.}. We can only obtain crude estimates of how many times each input frequency is actually detected. This can be done by looking at Fig.~\ref{fig:hist_all}, which represents the histograms of the 100 outputs for each method. These are upper limits to the rate of detection of the time-series simulator input frequencies, mostly because of the multiple peaks sometimes necessary to describe a single mode. 

 The two methods lead to very similar results. This indicates that the model we used is the main factor determining the outcome of the estimation process. Both methodologies are obviously very sensitive to the amplitude of the mode. Only in the 2800--3400 $\mu$Hz region do these algorithms find the input frequency $\sim$50\% of the time within a frequency interval corresponding to twice the natural resolution $\delta = 1/T$. This percentage, drops strongly below and above these values. 

In the MAP case, the main factor affecting the efficiency of the algorithm is the mode lifetime. More precisely, at higher frequencies, when the mode is most likely resolved, the proportion of correct detections decreases. It is in the range 60\%--100\% for frequencies between 1980~{\muhz} and $\sim$3400~{\muhz}. It drops significantly for eigenfrequencies above $\sim$3400 $\mu$Hz. In any case, the detection rates are much higher than those observed for CLEAN and {\tt SparSpec}. This is the reflection of the fact that we used different physical models to describe the signal. It is somewhat in disagreement with the findings of \citet{White10}.

\subsubsection{Impact of the prior formulation}\label{sect:impact}

To further understand how the priors affect the results, we tried to combined our CLEAN algorithm with constraints similar to those described by our prior on the frequencies. It should be noted that algorithms such as CLEAN have not been designed with a Bayesian perspective in mind \citep[see for instance][]{Schniter09}. Therefore, we could only try to mimic the impact of the prior. To this effect, we retained the idea that the frequency prior acts in analogy like a bandpass filter, which removes all signal outside the top-hat functions. We therefore applied such a filter (a sum of bandpass filters) to our spectrum. We then used the CLEAN algorithm to search only for two frequencies per individual bandpass filter. In a sense, this strategy is very similar to the {\textquotedblleft}ridge search{\textquotedblright} approach used for $\eta$ Boo by \citet{Kjeldsen95}. The corresponding output histogram is displayed in Figure~\ref{fig:hist_all}.
The information we get from this test is, of course, only qualitative, but it gives an interesting picture of the performance of the two algorithms under constraints that are fairly similar. 
 
We performed this test on the same 100 time-series sample. We can see that this definitely enhances the performance of the CLEAN algorithm. The detection rate increases everywhere, particularly in the low-frequency regions of the spectrum. In the high-frequency regions, the situation also improves, but the model is subject to limitations concerning the mode lifetimes, which perturb the estimation. However, the overall performance remains largely inferior to the outcome of the MAP strategy, which uses model (\ref{eq:sumf}). This is very revealing as to the effect the priors have on the final frequency estimates. It is often contended that Bayesian analysis may use too strong priors and retrieve only what as been defined in $\pi(\thetav)$ before the estimation. This is not entirely the case here. The priors on the frequencies we used are not so strong that any algorithm will be able to perform equally well under such a constraint. This result illustrates the subtle interplay between the numerical and statistical advantages of the Bayesian method mentioned in Sect.~\ref{sect:bayes}. Not only do the priors tighten the relevant volume in the space of parameters, but they also stabilize the fit to the data when using a more complex but also more accurate model, involving a larger number of parameters (higher dimension of the parameter space).

This is not the first time that Bayesian methods have been used on time series with such short time baseline \citep[][for instance]{Brewer07}. However, the very favourable case of {\18sco} allows us to contend that, provided the frequency priors are accurate enough, direct fits to the power spectrum are more efficient than classical time-domain modelling. A further step would be to test this claim with more sophisticated models for the spectrum \citep{Stahn08} or the time series \citep{Brewer09}.

\subsection{Large and small separations}\label{sect:separations}

\begin{figure}[t]
\center
\includegraphics[width=\columnwidth]{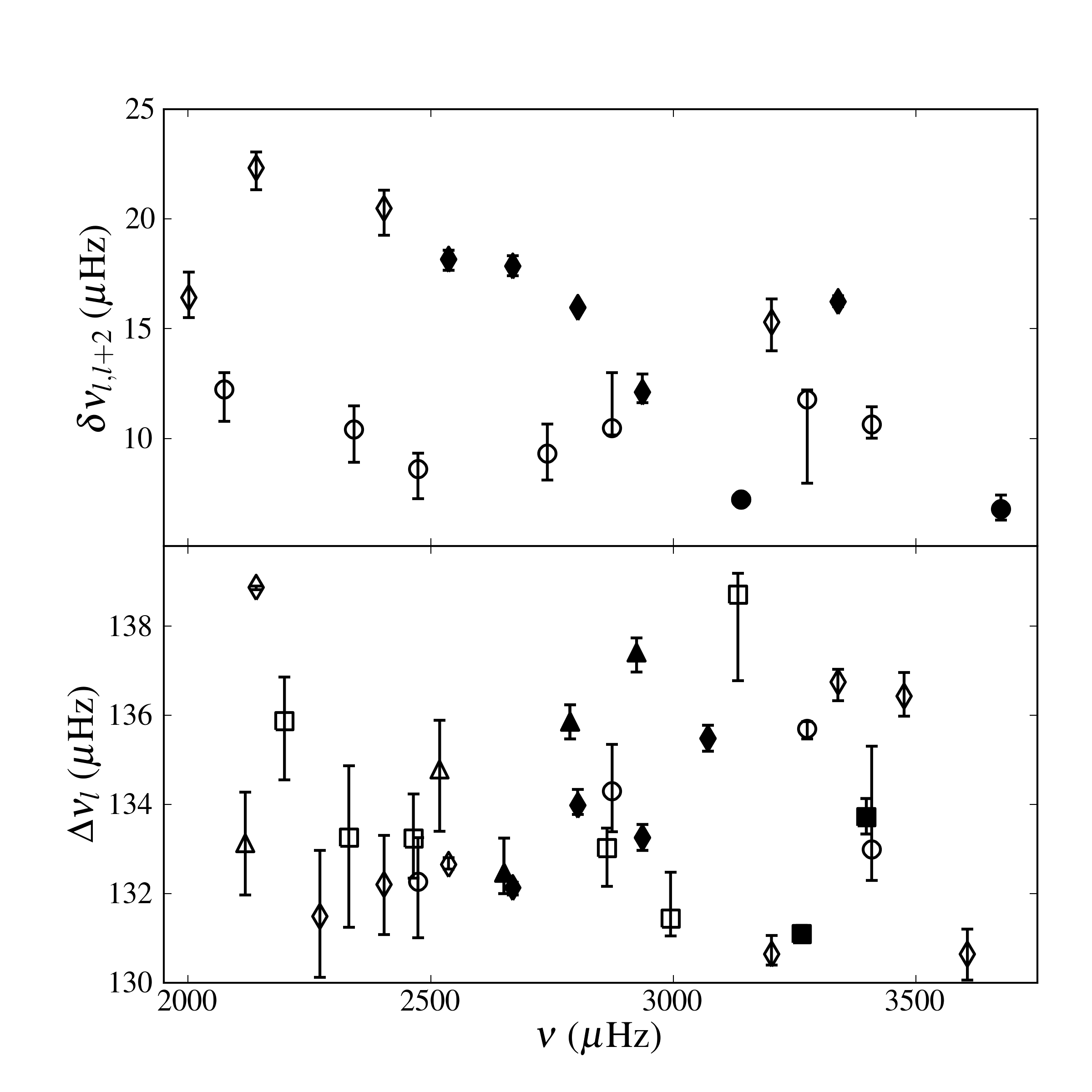}
\caption{Individual large (lower panel) and small (upper panel) separations for 18 Sco. The circles ($\circ$) corresponds to $l=0$, the diamonds ($\lozenge$) to $l=1$, the squares ($\square$) to $l=2$ and the triangles ($\triangle$) to $l=3$. Filled symbols mark the combinations using unflagged frequencies in Table~\ref{table:modes_mcmc}. A few error bars do not appear because they are smaller than their respective symbols.}
\label{fig:separations}
\end{figure}

Two common seismic indicators are the large and small separations, defined respectively by $\Delta\nu_{l}(n) = \nu_{n+1,l}-\nu_{n,l}$ and $\delta\nu_{l,l+2}(n) = \nu_{n,l} - \nu_{n-1,l+2}$. They both stem from a first analysis of the asymptotic relation for p modes. Their use has been popularized by the fact that they are supposed to be relatively free of the unknown surface effects affecting the oscillation frequencies. This however is only partially true and other combination of frequencies have been suggested in the literature \citep[e.g.,][]{Roxburgh03,Cunha07}. We nevertheless limit ourselves to these two quantities, which are plotted in Fig.~\ref{fig:separations}.

It is also interesting to compute the average values of the large and small separations. In order to estimate them, we use the MCMC samples. They are convenient to study densities of averages because, if we assume that the central-limit theorem roughly applies, we can expect to deal with Gaussian distribution (Benomar, private communication). This greatly simplifies the subsequent statistical analysis.

To estimate the average large separation we retained only the unflagged modes in Table~\ref{table:modes_mcmc}. Basing ourselves on the asymptotic relation, we consider that, for a fixed degree, the frequency is a linear function of the mode order, with slope the average large separation. We thus computed the derivative of $\nu(n)$ at each order for each degree and averaged over both quantities. The resulting distribution is well approximated by a Gaussian and we obtained $\langle \Delta \nu_{0,2}\rangle = 133.8\pm0.2$~$\mu$Hz. This  value agree with the estimate of Paper~I,  $\langle \Delta \nu_{0,2}\rangle = 134.4\pm0.3$~$\mu$Hz, at $2\sigma$ level.

This sheds a new light on the mass estimate we gave in Paper~I. Using these new values for the average large separation and the interferometric radius, one might evaluate the mass of 18 Sco to be $1.01\pm0.03$~M$_{\odot}$, which brings it even {\textquotedblleft}closer{\textquotedblright} to the Sun. Although this agrees within the 1$\sigma$ error bars with the value of Paper~I, it is relevant for modelling if one is to use on of these masses estimates in order to, for instance, apply a prior on the mass when modelling 18~Sco in the Bayesian framework \citep{Bazot08}. 

The average small separations are slightly more problematic. This is due to the fact that fewer values are available to average over, in particular for $\delta\nu_{02}$. We thus decided to include the frequencies from Table~\ref{table:modes_mcmc} flagged with a (\textasteriskcentered). Even though they display the multiple maxima, our idea is that the most prominent peaks in the distributions will be the main contributors to the density of the averaged value. We indeed found Gaussian distributions for $\langle \delta \nu_{0,2}\rangle$ and $\langle \delta \nu_{1,3}\rangle$ (the latter being much better approximated by such a distribution than the former). Using their first moments, we get $\langle \delta \nu_{0,2}\rangle = 9.4\pm 0.9$~$\mu$Hz and $\langle \delta \nu_{1,3}\rangle = 16.7\pm0.8$~$\mu$Hz. 

It is known that the small separations depend on frequency. However, to the first order, they can be approximated by constants $\delta \nu_{0,2}(n) \simeq 6D_0 \simeq 3\delta \nu_{1,3}/5$, with $D_0$ an integral containing the derivative of the sound speed \citep{Gough86,Gabriel89}.
The estimated values lead to a ratio $\langle \delta \nu_{0,2}\rangle/\langle \delta \nu_{1,3}\rangle = 0.57$, in good agreement with the theoretical expectations.

\section{Conclusion}

 We presented a detailed analysis of the ground-based seismic data obtained for the solar twin 18 Sco from the high-precision spectrograph HARPS. The sampling of the time series causes serious problems for stellar eigenfrequency estimation. We chose to use an MCMC algorithm in order to estimate the frequencies of 52 stellar pulsation modes. A careful examination of the posterior probability densities for each of them show that at least 21 are reliable and at least 19 others are worth consideration, even though the corresponding marginal PPDs are more difficult to analyze. 11 were rejected after comparison with a the MAP direct optimization methodology. We were able to estimate Bayesian credible intervals for these modes which reflect with some robustness the uncertainties of our data. 

By comparing with other estimation methods, we have discussed how reliable are the priors we used for the estimation. On the one hand, they are constraining enough to allow us to use a (relatively) realistic model. On the other hand, they are not so restrictive that they would impede a proper estimation. We note that our methodology can be further improved by increasing the efficiency of our MCMC algorithm (which would allow to relax further the priors on the parameters) and/or by using even more accurate models (which may require more conservative priors).

The individual eigenfrequencies obtained for 18~Sco allowed us to study some basic seismic estimators, including the large separations, whose estimation of the average value was addressed earlier in Paper~I. The two values agree at $2\sigma$ level, with the new one being lower. We derived a new value for the mass of the star slightly lower than the previous one. It remains to see how much this might affect the modelling of the star.

The sampling issues of our data certainly played an important part in producing the differences observed between the various methods used in our study. A next step would be to observe this star with more than one telescope, even though its magnitude makes such a task challenging for most of the ground-based instruments now available. Furthermore, we noticed that the length of the time series might imply that we only resolve a fraction of the detected modes. This could be resolved with a longer time basis.

\begin{acknowledgements}
The authors would like to thank the referee for his very careful report and careful comments. We feel that he helped to improve significantly this paper. This work was co-supported by grants SFRH/BPD/47994/2008, SFRH/BD/36240/2007 and PTDC/CTE-AST/098754/2008 from FCT/MCTES and FEDER, Portugal. MB thanks O.~Benomar for very interesting and fruitful discussions. A-MB and WJC acknowledge the support of the UK Science and Technology Facilities Council (STFC). Funding for the Stellar Astrophysics Centre is provided by The Danish National Research Foundation. The research is supported by the ASTERISK project (ASTERoseismic Investigations with SONG and Kepler) funded by the European Research Council (Grant agreement no.: 267864). 

\end{acknowledgements}

\bibliography{17963ref}

\begin{thebibliography}{73}
\expandafter\ifx\csname natexlab\endcsname\relax\def\natexlab#1{#1}\fi

\bibitem[{{Anderson} {et~al.}(1990){Anderson}, {Duvall}, \&
  {Jefferies}}]{Anderson90}
{Anderson}, E.~R., {Duvall}, Jr., T.~L., \& {Jefferies}, S.~M. 1990, \apj, 364,
  699

\bibitem[{{Appourchaux} {et~al.}(2008){Appourchaux}, {Michel}, {Auvergne},
  {Baglin}, {Toutain}, {Baudin}, {Benomar}, {Chaplin}, {Deheuvels}, {Samadi},
  {Verner}, {Boumier}, {Garc{\'{\i}}a}, {Mosser}, {Hulot}, {Ballot}, {Barban},
  {Elsworth}, {Jim{\'e}nez-Reyes}, {Kjeldsen}, {R{\'e}gulo}, \&
  {Roxburgh}}]{Appourchaux08}
{Appourchaux}, T., {Michel}, E., {Auvergne}, M., {et~al.} 2008, \aap, 488, 705

\bibitem[{{Appourchaux} {et~al.}(2009){Appourchaux}, {Samadi}, \&
  {Dupret}}]{Appourchaux09}
{Appourchaux}, T., {Samadi}, R., \& {Dupret}, M. 2009, \aap, 506, 1

\bibitem[{{Bazot} {et~al.}(2007){Bazot}, {Bouchy}, {Kjeldsen}, {Charpinet},
  {Laymand}, \& {Vauclair}}]{Bazot07}
{Bazot}, M., {Bouchy}, F., {Kjeldsen}, H., {et~al.} 2007, \aap, 470, 295

\bibitem[{{Bazot} {et~al.}(2008){Bazot}, {Bourguignon}, \&
  {Christensen-Dalsgaard}}]{Bazot08}
{Bazot}, M., {Bourguignon}, S., \& {Christensen-Dalsgaard}, J. 2008, Memorie
  della Societa Astronomica Italiana, 79, 660

\bibitem[{{Bazot} {et~al.}(2011){Bazot}, {Ireland}, {Huber}, {Bedding},
  {Broomhall}, {Campante}, {Carfantan}, {Chaplin}, {Elsworth}, {Mel{\'e}ndez},
  {Petit}, {Th{\'e}ado}, {van Grootel}, {Arentoft}, {Asplund}, {Castro},
  {Christensen-Dalsgaard}, {Do Nascimento}, {Dintrans}, {Dumusque}, {Kjeldsen},
  {McAlister}, {Metcalfe}, {Monteiro}, {Santos}, {Sousa}, {Sturmann},
  {Sturmann}, {Ten Brummelaar}, {Turner}, \& {Vauclair}}]{Bazot11}
{Bazot}, M., {Ireland}, M.~J., {Huber}, D., {et~al.} 2011, \aap, 526, L4

\bibitem[{{Bazot} {et~al.}(2005){Bazot}, {Vauclair}, {Bouchy}, \&
  {Santos}}]{Bazot05}
{Bazot}, M., {Vauclair}, S., {Bouchy}, F., \& {Santos}, N.~C. 2005, \aap, 440,
  615

\bibitem[{{Bedding} \& {Kjeldsen}(2010)}]{Bedding10b}
{Bedding}, T.~R. \& {Kjeldsen}, H. 2010, Communications in Asteroseismology,
  161, 3

\bibitem[{{Bedding} {et~al.}(2010){Bedding}, {Kjeldsen}, {Campante},
  {Appourchaux}, {Bonanno}, {Chaplin}, {Garcia}, {Marti{\'c}}, {Mosser},
  {Butler}, {Bruntt}, {Kiss}, {O'Toole}, {Kambe}, {Ando}, {Izumiura}, {Sato},
  {Hartmann}, {Hatzes}, {Barban}, {Berthomieu}, {Michel}, {Provost},
  {Turck-Chi{\`e}ze}, {Lebrun}, {Schmitt}, {Bertaux}, {Benatti}, {Claudi},
  {Cosentino}, {Leccia}, {Frandsen}, {Brogaard}, {Glowienka}, {Grundahl},
  {Stempels}, {Arentoft}, {Bazot}, {Christensen-Dalsgaard}, {Dall}, {Karoff},
  {Lundgreen-Nielsen}, {Carrier}, {Eggenberger}, {Sosnowska}, {Wittenmyer},
  {Endl}, {Metcalfe}, {Hekker}, \& {Reffert}}]{Bedding10}
{Bedding}, T.~R., {Kjeldsen}, H., {Campante}, T.~L., {et~al.} 2010, \apj, 713,
  935

\bibitem[{{Benomar} {et~al.}(2009){Benomar}, {Appourchaux}, \&
  {Baudin}}]{Benomar09}
{Benomar}, O., {Appourchaux}, T., \& {Baudin}, F. 2009, \aap, 506, 15

\bibitem[{{Bouchy} {et~al.}(2005){Bouchy}, {Bazot}, {Santos}, {Vauclair}, \&
  {Sosnowska}}]{Bouchy05}
{Bouchy}, F., {Bazot}, M., {Santos}, N.~C., {Vauclair}, S., \& {Sosnowska}, D.
  2005, \aap, 440, 609

\bibitem[{{Bouchy} \& {Carrier}(2001)}]{Bouchy01}
{Bouchy}, F. \& {Carrier}, F. 2001, \aap, 374, L5

\bibitem[{{Bourguignon} {et~al.}(2007){Bourguignon}, {Carfantan}, \&
  {B{\"o}hm}}]{Bourgui07}
{Bourguignon}, S., {Carfantan}, H., \& {B{\"o}hm}, T. 2007, \aap, 462, 379

\bibitem[{{Brand{\~a}o} {et~al.}(2011){Brand{\~a}o}, {Do{\u g}an},
  {Christensen-Dalsgaard}, {Cunha}, {Bedding}, {Metcalfe}, {Kjeldsen},
  {Bruntt}, \& {Arentoft}}]{Brandao11}
{Brand{\~a}o}, I.~M., {Do{\u g}an}, G., {Christensen-Dalsgaard}, J., {et~al.}
  2011, \aap, 527, A37+

\bibitem[{{Bretthorst}(2000)}]{Bretthorst00}
{Bretthorst}, G.~L. 2000, in Bulletin of the American Astronomical Society,
  Vol.~32, Bulletin of the American Astronomical Society, 1438

\bibitem[{{Brewer} {et~al.}(2007){Brewer}, {Bedding}, {Kjeldsen}, \&
  {Stello}}]{Brewer07}
{Brewer}, B.~J., {Bedding}, T.~R., {Kjeldsen}, H., \& {Stello}, D. 2007, \apj,
  654, 551

\bibitem[{{Brewer} \& {Stello}(2009)}]{Brewer09}
{Brewer}, B.~J. \& {Stello}, D. 2009, \mnras, 395, 2226

\bibitem[{{Broomhall} {et~al.}(2009){Broomhall}, {Chaplin}, {Davies},
  {Elsworth}, {Fletcher}, {Hale}, {Miller}, \& {New}}]{Broomhall09}
{Broomhall}, A., {Chaplin}, W.~J., {Davies}, G.~R., {et~al.} 2009, \mnras, 396,
  L100

\bibitem[{{Campante} {et~al.}(2011){Campante}, {Handberg}, {Mathur},
  {Appourchaux}, {Bedding}, {Chaplin}, {Garc{\'{\i}}a}, {Mosser}, {Benomar},
  {Bonanno}, {Corsaro}, {Fletcher}, {Gaulme}, {Hekker}, {Karoff}, {R{\'e}gulo},
  {Salabert}, {Verner}, {White}, {Houdek}, {Brand{\~a}o}, {Creevey}, {Do{\v
  g}an}, {Bazot}, {Christensen-Dalsgaard}, {Cunha}, {Elsworth}, {Huber},
  {Kjeldsen}, {Lundkvist}, {Molenda-{\.Z}akowicz}, {Monteiro}, {Stello},
  {Clarke}, {Girouard}, \& {Hall}}]{Campante11}
{Campante}, T.~L., {Handberg}, R., {Mathur}, S., {et~al.} 2011, \aap, 534, A6

\bibitem[{{Carrier} \& {Bourban}(2003)}]{Carrier03}
{Carrier}, F. \& {Bourban}, G. 2003, \aap, 406, L23

\bibitem[{{Cayrel de Strobel} {et~al.}(1981){Cayrel de Strobel}, {Knowles},
  {Hernandez}, \& {Bentolila}}]{CdS81}
{Cayrel de Strobel}, G., {Knowles}, N., {Hernandez}, G., \& {Bentolila}, C.
  1981, \aap, 94, 1

\bibitem[{{Chaplin} {et~al.}(2002){Chaplin}, {Elsworth}, {Isaak}, {Marchenkov},
  {Miller}, {New}, {Pinter}, \& {Appourchaux}}]{Chaplin02}
{Chaplin}, W.~J., {Elsworth}, Y., {Isaak}, G.~R., {et~al.} 2002, \mnras, 336,
  979

\bibitem[{{Chaplin} {et~al.}(2008){Chaplin}, {Houdek}, {Appourchaux},
  {Elsworth}, {New}, \& {Toutain}}]{Chaplin08}
{Chaplin}, W.~J., {Houdek}, G., {Appourchaux}, T., {et~al.} 2008, \aap, 485,
  813

\bibitem[{{Chaplin} {et~al.}(2005){Chaplin}, {Houdek}, {Elsworth}, {Gough},
  {Isaak}, \& {New}}]{Chaplin05}
{Chaplin}, W.~J., {Houdek}, G., {Elsworth}, Y., {et~al.} 2005, \mnras, 360, 859

\bibitem[{{Cunha} \& {Metcalfe}(2007)}]{Cunha07}
{Cunha}, M.~S. \& {Metcalfe}, T.~S. 2007, \apj, 666, 413

\bibitem[{{De Rosa} {et~al.}(2000){De Rosa}, {Duvall}, \& {Toomre}}]{DeRosa00}
{De Rosa}, M., {Duvall}, Jr., T.~L., \& {Toomre}, J. 2000, \solphys, 192, 351

\bibitem[{{Deeming}(1975)}]{Deeming75}
{Deeming}, T.~J. 1975, \apss, 36, 137

\bibitem[{{Deheuvels} \& {Michel}(2010)}]{Deheuvels10}
{Deheuvels}, S. \& {Michel}, E. 2010, \apss, 328, 259

\bibitem[{{Do{\u g}an} {et~al.}(2010){Do{\u g}an}, {Brand{\~a}o}, {Bedding},
  {Christensen-Dalsgaard}, {Cunha}, \& {Kjeldsen}}]{Dogan10}
{Do{\u g}an}, G., {Brand{\~a}o}, I.~M., {Bedding}, T.~R., {et~al.} 2010, \apss,
  328, 101

\bibitem[{{Dumusque} {et~al.}(2011){Dumusque}, {Udry}, {Lovis}, {Santos}, \&
  {Monteiro}}]{Dumusque11}
{Dumusque}, X., {Udry}, S., {Lovis}, C., {Santos}, N.~C., \& {Monteiro},
  M.~J.~P.~F.~G. 2011, \aap, 525, A140

\bibitem[{{Fletcher} {et~al.}(2009){Fletcher}, {Chaplin}, {Elsworth}, \&
  {New}}]{Fletcher09}
{Fletcher}, S.~T., {Chaplin}, W.~J., {Elsworth}, Y., \& {New}, R. 2009, \apj,
  694, 144

\bibitem[{{Foglizzo}(1998)}]{Foglizzo98b}
{Foglizzo}, T. 1998, \aap, 339, 261

\bibitem[{{Foglizzo} {et~al.}(1998){Foglizzo}, {Garcia}, {Boumier}, {Charra},
  {Gabriel}, {Grec}, {Robillot}, {Roca Cortes}, {Turck-Chieze}, \&
  {Ulrich}}]{Foglizzo98a}
{Foglizzo}, T., {Garcia}, R.~A., {Boumier}, P., {et~al.} 1998, \aap, 330, 341

\bibitem[{{Gabriel}(1989)}]{Gabriel89}
{Gabriel}, M. 1989, \aap, 226, 278

\bibitem[{{Gaulme} {et~al.}(2009){Gaulme}, {Appourchaux}, \&
  {Boumier}}]{Gaulme09}
{Gaulme}, P., {Appourchaux}, T., \& {Boumier}, P. 2009, \aap, 506, 7

\bibitem[{{Gough}(1986)}]{Gough86}
{Gough}, D.~O. 1986, in Hydrodynamic and Magnetodynamic Problems in the Sun and
  Stars, ed. {Y.~Osaki}, 117

\bibitem[{{Gray} \& {Desikachary}(1973)}]{Gray73}
{Gray}, D.~F. \& {Desikachary}, K. 1973, \apj, 181, 523

\bibitem[{{Gregory}(2005)}]{Gregory05}
{Gregory}, P.~C. 2005, {Bayesian Logical Data Analysis for the Physical
  Sciences: A Comparative Approach with `Mathematica' Support} (Cambridge
  University Press)

\bibitem[{{Guenther} {et~al.}(2008){Guenther}, {Kallinger}, {Gruberbauer},
  {Huber}, {Weiss}, {Kuschnig}, {Demarque}, {Robinson}, {Matthews}, {Moffat},
  {Rucinski}, {Sasselov}, \& {Walker}}]{Guenther08}
{Guenther}, D.~B., {Kallinger}, T., {Gruberbauer}, M., {et~al.} 2008, \apj,
  687, 1448

\bibitem[{{Gustafsson}(1998)}]{Gustafsson98}
{Gustafsson}, B. 1998, Space Science Reviews, 85, 419

\bibitem[{{Gustafsson}(2008)}]{Gustafsson08}
{Gustafsson}, B. 2008, Physica Scripta Volume T, 130, 014036

\bibitem[{{Handberg} \& {Campante}(2011)}]{Handberg11}
{Handberg}, R. \& {Campante}, T.~L. 2011, \aap, 527, A56

\bibitem[{{Harvey}(1985)}]{Harvey85}
{Harvey}, J. 1985, in ESA Special Publication, Vol. 235, Future Missions in
  Solar, Heliospheric \& Space Plasma Physics, ed. {E.~Rolfe \& B.~Battrick},
  199--208

\bibitem[{{Hastings}(1970)}]{Hastings70}
{Hastings}, W.~K. 1970, Biometrika, 57, 97

\bibitem[{{Houdek} {et~al.}(1999){Houdek}, {Balmforth},
  {Christensen-Dalsgaard}, \& {Gough}}]{Houdek99}
{Houdek}, G., {Balmforth}, N.~J., {Christensen-Dalsgaard}, J., \& {Gough},
  D.~O. 1999, \aap, 351, 582

\bibitem[{{Jim{\'e}nez-Reyes} {et~al.}(2008){Jim{\'e}nez-Reyes}, {Chaplin},
  {Garc{\'{\i}}a}, {Appourchaux}, {Baudin}, {Boumier}, {Elsworth}, {Fletcher},
  {Lazrek}, {Leibacher}, {Lochard}, {New}, {R{\'e}gulo}, {Salabert}, {Toutain},
  {Verner}, \& {Wachter}}]{JR08}
{Jim{\'e}nez-Reyes}, S.~J., {Chaplin}, W.~J., {Garc{\'{\i}}a}, R.~A., {et~al.}
  2008, \mnras, 389, 1780

\bibitem[{{Kjeldsen} {et~al.}(2008){Kjeldsen}, {Bedding}, {Arentoft}, {Butler},
  {Dall}, {Karoff}, {Kiss}, {Tinney}, \& {Chaplin}}]{Kjeldsen08}
{Kjeldsen}, H., {Bedding}, T.~R., {Arentoft}, T., {et~al.} 2008, \apj, 682,
  1370

\bibitem[{{Kjeldsen} {et~al.}(2005){Kjeldsen}, {Bedding}, {Butler},
  {Christensen-Dalsgaard}, {Kiss}, {McCarthy}, {Marcy}, {Tinney}, \&
  {Wright}}]{Kjeldsen05}
{Kjeldsen}, H., {Bedding}, T.~R., {Butler}, R.~P., {et~al.} 2005, \apj, 635,
  1281

\bibitem[{{Kjeldsen} {et~al.}(1995){Kjeldsen}, {Bedding}, {Viskum}, \&
  {Frandsen}}]{Kjeldsen95}
{Kjeldsen}, H., {Bedding}, T.~R., {Viskum}, M., \& {Frandsen}, S. 1995, \aj,
  109, 1313

\bibitem[{{Lefebvre} {et~al.}(2008){Lefebvre}, {Garc{\'{\i}}a},
  {Jim{\'e}nez-Reyes}, {Turck-Chi{\`e}ze}, \& {Mathur}}]{Lefebvre08}
{Lefebvre}, S., {Garc{\'{\i}}a}, R.~A., {Jim{\'e}nez-Reyes}, S.~J.,
  {Turck-Chi{\`e}ze}, S., \& {Mathur}, S. 2008, \aap, 490, 1143

\bibitem[{{Lomb}(1976)}]{Lomb76}
{Lomb}, N.~R. 1976, \apss, 39, 447

\bibitem[{{Marti{\'c}} {et~al.}(2004){Marti{\'c}}, {Lebrun}, {Appourchaux}, \&
  {Korzennik}}]{Martic04}
{Marti{\'c}}, M., {Lebrun}, J., {Appourchaux}, T., \& {Korzennik}, S.~G. 2004,
  \aap, 418, 295

\bibitem[{{Mathur} {et~al.}(2010){Mathur}, {Garc{\'{\i}}a}, {Catala}, {Bruntt},
  {Mosser}, {Appourchaux}, {Ballot}, {Creevey}, {Gaulme}, {Hekker}, {Huber},
  {Karoff}, {Piau}, {R{\'e}gulo}, {Roxburgh}, {Salabert}, {Verner}, {Auvergne},
  {Baglin}, {Chaplin}, {Elsworth}, {Michel}, {Samadi}, {Sato}, \&
  {Stello}}]{Mathur10}
{Mathur}, S., {Garc{\'{\i}}a}, R.~A., {Catala}, C., {et~al.} 2010, \aap, 518,
  A53

\bibitem[{{Mel{\'e}ndez} \& {Ram{\'{\i}}rez}(2007)}]{Melendez07}
{Mel{\'e}ndez}, J. \& {Ram{\'{\i}}rez}, I. 2007, \apjl, 669, L89

\bibitem[{{Mel{\'e}ndez} {et~al.}(2010){Mel{\'e}ndez}, {Schuster}, {Silva},
  {Ram{\'{\i}}rez}, {Casagrande}, \& {Coelho}}]{Melendez10}
{Mel{\'e}ndez}, J., {Schuster}, W.~J., {Silva}, J.~S., {et~al.} 2010, \aap,
  522, A98

\bibitem[{{Metcalfe} {et~al.}(2010){Metcalfe}, {Monteiro}, {Thompson},
  {Molenda-{\.Z}akowicz}, {Appourchaux}, {Chaplin}, {Do{\u g}an},
  {Eggenberger}, {Bedding}, {Bruntt}, {Creevey}, {Quirion}, {Stello},
  {Bonanno}, {Silva Aguirre}, {Basu}, {Esch}, {Gai}, {Di Mauro}, {Kosovichev},
  {Kitiashvili}, {Su{\'a}rez}, {Moya}, {Piau}, {Garc{\'{\i}}a}, {Marques},
  {Frasca}, {Biazzo}, {Sousa}, {Dreizler}, {Bazot}, {Karoff}, {Frandsen},
  {Wilson}, {Brown}, {Christensen-Dalsgaard}, {Gilliland}, {Kjeldsen},
  {Campante}, {Fletcher}, {Handberg}, {R{\'e}gulo}, {Salabert}, {Schou},
  {Verner}, {Ballot}, {Broomhall}, {Elsworth}, {Hekker}, {Huber}, {Mathur},
  {New}, {Roxburgh}, {Sato}, {White}, {Borucki}, {Koch}, \&
  {Jenkins}}]{Metcalfe10}
{Metcalfe}, T.~S., {Monteiro}, M.~J.~P.~F.~G., {Thompson}, M.~J., {et~al.}
  2010, \apj, 723, 1583

\bibitem[{{Metropolis}(1953)}]{Metropolis53}
{Metropolis}, N. 1953, \jcp, 21, 1087

\bibitem[{{Miglio} \& {Montalb{\'a}n}(2005)}]{Miglio05}
{Miglio}, A. \& {Montalb{\'a}n}, J. 2005, \aap, 441, 615

\bibitem[{{Nordlund} {et~al.}(2009){Nordlund}, {Stein}, \&
  {Asplund}}]{Nordlund09}
{Nordlund}, {\AA}., {Stein}, R.~F., \& {Asplund}, M. 2009, Living Reviews in
  Solar Physics, 6, 2

\bibitem[{{Petit} {et~al.}(2008){Petit}, {Dintrans}, {Solanki}, {Donati},
  {Auri{\`e}re}, {Ligni{\`e}res}, {Morin}, {Paletou}, {Ramirez}, {Catala}, \&
  {Fares}}]{Petit08}
{Petit}, P., {Dintrans}, B., {Solanki}, S.~K., {et~al.} 2008, \mnras, 388, 80

\bibitem[{Robert(2007)}]{Robert07}
Robert, C.~P. 2007, {The Bayesian Choice: From Decision-Theoretic Foundations
  to Computational Implementation (Springer Texts in Statistics)}, 2nd edn.
  (Springer Verlag, New York)

\bibitem[{Robert \& Casella(1999)}]{Robert99}
Robert, C.~P. \& Casella, G. 1999, {Monte Carlo Statistical Methods}, 1st edn.
  (Springer-Verlag)

\bibitem[{{Roberts} {et~al.}(1987){Roberts}, {Lehar}, \& {Dreher}}]{Roberts87}
{Roberts}, D.~H., {Lehar}, J., \& {Dreher}, J.~W. 1987, \aj, 93, 968

\bibitem[{{Roxburgh} \& {Vorontsov}(2003)}]{Roxburgh03}
{Roxburgh}, I.~W. \& {Vorontsov}, S.~V. 2003, \aap, 411, 215

\bibitem[{{Scargle}(1982)}]{Scargle82}
{Scargle}, J.~D. 1982, \apj, 263, 835

\bibitem[{Schniter {et~al.}(2009)Schniter, Potter, \& Ziniel}]{Schniter09}
Schniter, P., Potter, L.~C., \& Ziniel, J. 2009, Fast Bayesian Matching
  Pursuit: Model Uncertainty and Parameter Estimation for Sparse Linear Models

\bibitem[{{Shine} {et~al.}(2000){Shine}, {Simon}, \& {Hurlburt}}]{Shine00}
{Shine}, R.~A., {Simon}, G.~W., \& {Hurlburt}, N.~E. 2000, \solphys, 193, 313

\bibitem[{{Stahn} \& {Gizon}(2008)}]{Stahn08}
{Stahn}, T. \& {Gizon}, L. 2008, \solphys, 251, 31

\bibitem[{{Tassoul}(1980)}]{Tassoul80}
{Tassoul}, M. 1980, \apjs, 43, 469

\bibitem[{{Title} {et~al.}(1989){Title}, {Tarbell}, {Topka}, {Ferguson},
  {Shine}, \& {SOUP Team}}]{Title89}
{Title}, A.~M., {Tarbell}, T.~D., {Topka}, K.~P., {et~al.} 1989, \apj, 336, 475

\bibitem[{{Vandakurov}(1967)}]{Vandakurov67}
{Vandakurov}, Y.~V. 1967, \azh, 44, 786

\bibitem[{{White} {et~al.}(2010){White}, {Brewer}, {Bedding}, {Stello}, \&
  {Kjeldsen}}]{White10}
{White}, T.~R., {Brewer}, B.~J., {Bedding}, T.~R., {Stello}, D., \& {Kjeldsen},
  H. 2010, Communications in Asteroseismology, 161, 39

\bibitem[{{Zechmeister} \& {K{\"u}rster}(2009)}]{Zechmeister09}
{Zechmeister}, M. \& {K{\"u}rster}, M. 2009, \aap, 496, 577

\end{thebibliography}

\end{document}